\documentclass[11pt,titlepage]{article}
\usepackage[round,numbers,sort&compress]{natbib}
\usepackage{epsfig,amsmath,amssymb,amsfonts,bm,subfigure,setspace,graphicx}
\usepackage{authblk}

\newcommand{\p}{\partial}
\newcommand{\ds}{\displaystyle}
\newcommand{\bea}{\begin{eqnarray}}
\newcommand{\eea}{\end{eqnarray}}
\newcommand{\nn}{\nonumber}
\renewcommand{\(}{\left(}
\renewcommand{\)}{\right)}

\newcommand{\bc}{\begin{center}}
\newcommand{\ec}{\end{center}}

\title{Facilitated diffusion framework for transcription factor search with conformational changes}
\author[*]{J\'er\^ome Cartailler}
\author[**]{J\"urgen Reingruber\thanks{ Corresponding author: reingrub@ens.fr} }
\affil[*]{Ecole Normale Sup\'erieure, 46 rue d’Ulm, 75005 Paris , France.}
\affil[**]{INSERM U1024, Paris, France.}
\date{}
\begin{document}

\maketitle

\abstract{
Cellular responses often require the fast activation or repression of specific genes, which depends on Transcription Factors (TFs) that have to quickly find the promoters of these genes within a large genome. Transcription Factors (TFs) search for their DNA promoter target by alternating between bulk diffusion and sliding along the DNA, a mechanism known as facilitated diffusion. We study a  facilitated diffusion framework with switching between three search modes: a bulk mode and two sliding modes triggered by conformational changes between two protein conformations. In one conformation (search mode) the TF interacts unspecifically with the DNA backbone resulting in fast sliding. In the other conformation (recognition mode) it interacts specifically and strongly with DNA base pairs leading to slow displacement. From the bulk, a TF associates with the DNA at a random position that is correlated with the previous dissociation point, which implicitly is a function of the DNA structure. The target affinity depends on the conformation. We derive exact expressions for the mean first passage time (MFPT) to bind to the promoter and the conditional probability to bind before detaching when arriving at the promoter site. We systematically explore the parameter space and compare various search scenarios. We compare our results with experimental data for the dimeric Lac repressor search in E.Coli bacteria. We find that a coiled DNA conformation is absolutely necessary for a fast MFPT. With frequent spontaneous conformational changes, a fast search time is achieved even when a TF becomes immobilized in the recognition state due to the specific bindings. We find a MFPT compatible with experimental data in presence of a specific TF-DNA interaction energy that has a Gaussian distribution with a large variance.

}

\section*{Introduction}
Transcription factors (TFs) regulate gene activation by binding to DNA promoter sites. To enable a fast cellular response that relies on the activation or repression of specific genes, TFs perform a facilitated diffusion search where they alternate between three dimensional (3D) diffusion in the bulk and 1D diffusion (sliding) along the DNA (for reviews see \cite{ZabetAdryan_Review2012,Kolomeisky_TfsearchReview_2011,Tafvizietal_Review2011,LiElf_Review2007,Halford_Review_NuclARes2004,HippelBerg_JBC1989}). Initially, facilitated diffusion was introduced to explain the experimental finding that the in-vitro association rate of the Lac-I repressor with its promoter sites placed on $\lambda$-phage DNA was around 100 times larger than the Smoluchowski limit $\sim 10^{8}\,M^{-1}s^{-1}$ for a 3D diffusion process \cite{Riggs_JMolBiol1970}. Theoretical considerations showed that a search that alternates between 3D diffusion and 1D sliding can have a higher association rate compared to a pure 3D search \cite{RichterEigen_BiophysChem1974,BergBlomberg_ChemPhys1976,WinterBergvHippel_1_Biochem1981}.
With dilute DNA the search time is dominated by the 3D excursions between subsequent DNA binding events, and sliding increases the association rate by enlarging the effective target size (antenna effect). Later on single molecule techniques provided a direct experimental proof of the facilitated diffusion mechanism  \cite{Hammaretal_Science2012,Tafvizietal_P53SingleMolecule_PNAS2011,Elf_Science2007,WangAustinCox_PRL2006}.

It has also soon been realized \cite{WinterBergvHippel_2_Biochem1981,BergBlomberg_ChemPhys1976} that frequent bindings to the DNA are problematic because sliding along the DNA is slow due to strong TF-DNA interactions \cite{Elf_Science2007,WangAustinCox_PRL2006}. In a dense DNA environment with frequent bindings to the DNA and slow 1D diffusion, the antenna effect becomes negligible and facilitated diffusion is slower compared to a pure 3D search. For example, in E.Coli with a volume $|V|\sim 1\mu m^3$, the measured search time of the Lac repressor for its promoter site is $\tau \sim 350s$ \cite{Elf_Science2007,Hammaretal_Science2012}. This corresponds to an association rate $k_{a}=N_{Av}V/\tau \sim 10^6 M^{-1} s^{-1}$, much lower than the Smoluchowski limit. If a TF could specifically bind only to its promoter site and bounce off from the rest of the DNA, the search time would be extremely fast around $\sim {V}/({4\pi DR})\sim 5s$. However, because a TF cannot already recognize its target from the bulk, frequent DNA associations are essential. Thus, the question arises: How is a fast search possible within a large genome despite of facilitated diffusion ?

When a TF is bound to the DNA and interacts with the underlying base pairs (bps), the diffusion coefficient for sliding decays exponentially with the variance of the binding energy distribution \cite{SlutskyKardarMirny_PRE2004,ZwanzigPNAS1988}. With a simple facilitated diffusion model that comprises sliding along the DNA and uniform redistributions in 3D, one finds that a search time of the order of minutes is only compatible with a variance $\lesssim 1.5k_B T$ \cite{SlutskyMirny_BJ2004}. In contrast, binding energy estimates for the {\it Cro} and {\it PurR} TF reveal a much larger variance around $5-6 k_BT$ \cite{SlutskyMirny_BJ2004,Gerlandetal_TFinteraction_PNAS2002}. This indicates that a simple diffusion process is not sufficient to explain the search dynamics when a TF is bound to the DNA. It has been proposed that a TF switches between two protein conformations with different binding affinities to the DNA \cite{WinterBergvHippel_2_Biochem1981,SlutskyMirny_BJ2004}. In the search conformation, a TF interacts only non-specifically with the DNA backbone leading to a smooth energy profile and fast diffusion. In the recognition conformation, a TF interacts specifically with the underlying DNA sequence resulting in a rough energy landscape and slow diffusion. Conformational changes of the TF protein are indeed supported by experimental observations \cite{Leithetal_P53SlidingKinetics_PNAS2012,Tafvizietal_P53SingleMolecule_PNAS2011,Kalodimos_Science2004,Hortonetal_NonspecificSpecificMethyltransferase_Cell2005,ViadiuAggarwal_ModelForSliding_MolCell2000,Albrightetal_NonspecificSpecificCro_ProteinSci1998}.

\begin{figure}[t]
\begin{center}
\includegraphics[scale=0.23]{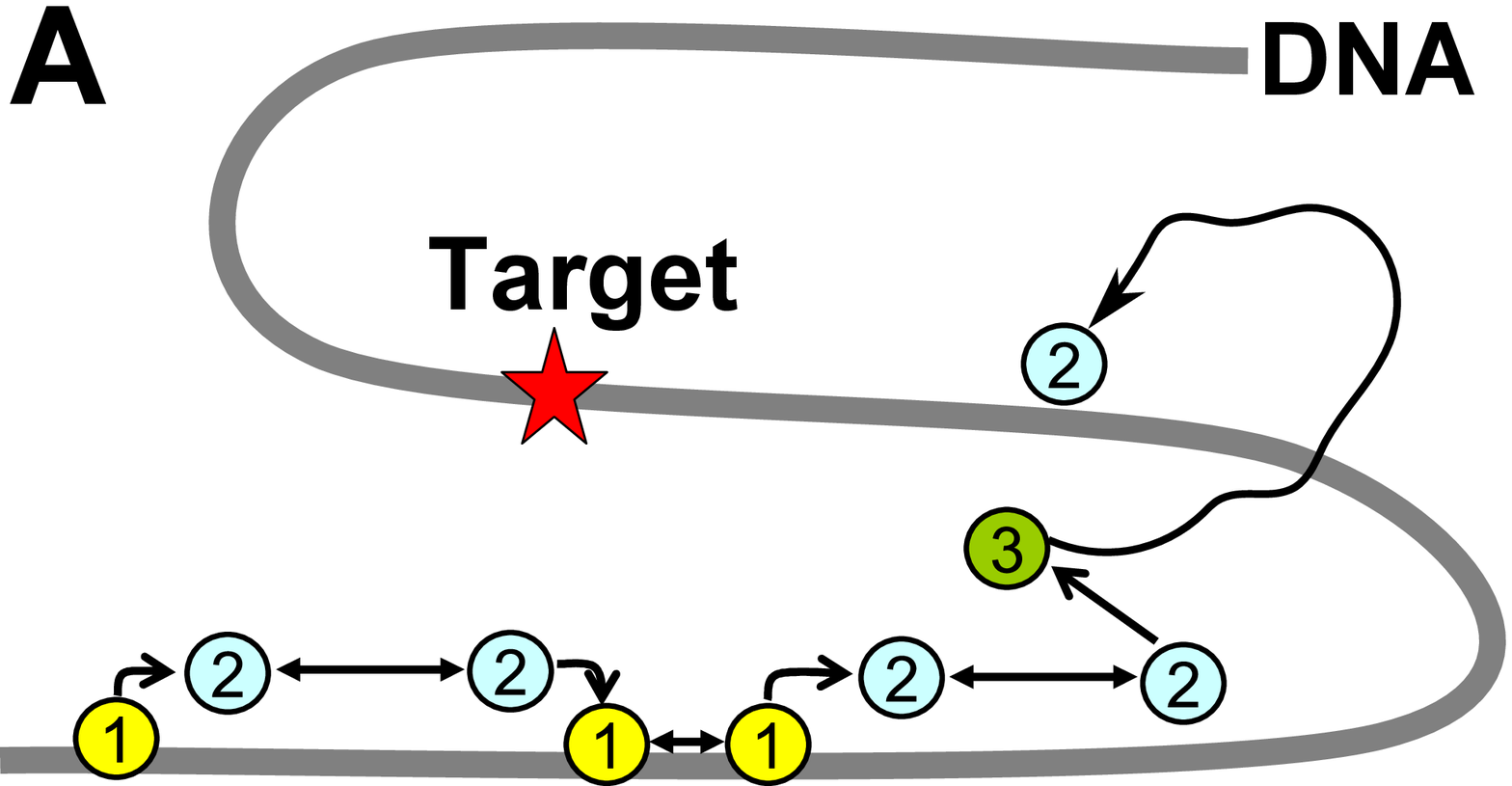} \hspace{0.7 cm}
\includegraphics[scale=0.18]{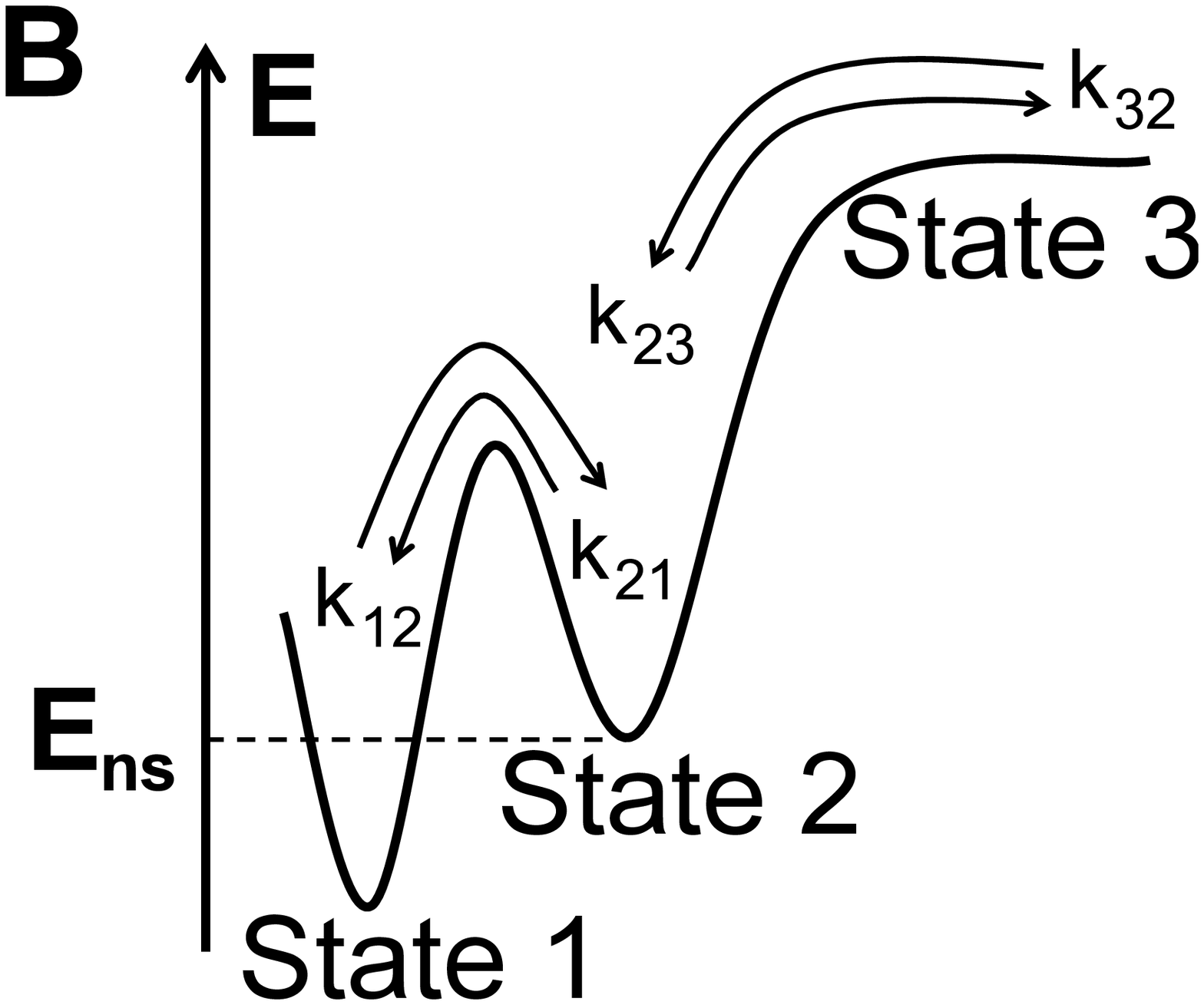}
      \caption{{\bf Facilitated diffusion framework.} (A) Schematic of the search process: In state 3 a TF is freely diffusing in the bulk. It attaches to the DNA at a random position following a Gaussian distribution centered around the previous detaching position. In state 1 and 2 a TF is attached to the DNA and diffuses  along the DNA. In state 1 it specifically interacts with the underlying DNA sequence and diffusion is slow. In state 2 it non-specifically interacts with the DNA backbone and diffusion is fast. The binding affinity to the target depends on the state. (B) Switchings between states occurs with Poissonian rates. The rate $k_{12}$ depends on the energy of the specific interaction.}
       \label{fig_scenario}
\end{center}
\end{figure}

In this work we investigate a general framework for a facilitated diffusion search with conformational changes. We analytically derive the mean first passage time (MFPT) to bind to the target and the conditional probability to bind before dissociation when a TF arrives at the target site. We further compute the ratio of the time spent in the bulk compared to attached to the DNA, the apparent diffusion constant for sliding along the DNA, and the average sliding distance before detaching. We consider a search process with Poissonian switchings between three states (Fig.~\ref{fig_scenario}). State 1 and 2 (recognition and search mode) correspond to two different protein conformations with conformation dependent TF-DNA interactions. Therefore the diffusion coefficients for sliding along the DNA and the target affinity depend on the conformation. In state 3, a TF is diffusing in the bulk and it associates with the DNA at a random position following a Gaussian distribution centered around the previous dissociation point. By modifying the target affinity in state 2 we evaluate the impact of induced switchings at the target site and the effect that a TF misses the target when arriving at the target site. By varying the correlation distance between dissociation and association point we estimate how the DNA conformation and coiling affect the MFPT. With our analytic expressions we can precisely evaluate the whole parameter space. We analyze various search scenarios within the same framework, which is important to accurately compare results. Other approaches partly rely on MFPT analysis, kinetic theory, thermodynamic equilibrium considerations, scaling arguments and other approximations, which complicates a comparison of results obtained with different methods and approximations  \cite{MahmutovicEtal_NuclAcidRes2015,Zhou_SwitchinTFsearch_PNAS2011,Murugan_TFSearch_BJ2010,Diazdelarosa_BJ2010,Mirnyetal_JPhysA2009,HuGrosbergBruinsma_BJ2008,Gerlandetal_TFinteraction_PNAS2002}. A clean MFPT analysis for a 3 states switching model with maximal target affinity in state 1 and no affinity in state 2 has been performed in \cite{BauerMetzler_BJ2012}. Compared to \cite{ReingruberHolcman_PRE2011}, the authors additionally consider 3D excursions using a closed-cell approach as described in \cite{BergBlomberg_ChemPhys1976}. However, due to the difficulty to compute the 3D kernel, only the asymptotic limits corresponding to uniform redistributions and a rod-like DNA are discussed. The first passage time distribution for a switching process between two 1D states has been studied in \cite{HuGrosbergBruinsma_PhysA2009}.

\section*{Model}

\subsection*{Model description and MFPT analysis}
We start by presenting the mathematical framework and the MFPT analysis. We postpone the description of the biological motivation to the results part. We consider a search that switches between 3 states with Poissonian switching rates $k_{ij}$ (Fig.~\ref{fig_scenario}). In state 1 and 2 a TF is attached to the DNA of length $2L$ and slides with state dependent diffusion constants $D_1$ and $D_2$. To simplify the analysis we consider that the target is located at the center. An off-centered target results in a higher MFPT up to maximally a factor of 4 if the target is located at the  periphery (assuming that the MFPT scales $\sim L^2$)   \cite{VekslerKolomeisky_JPhysChemB2013,CoppeyBenichou_TFSearch_BJ2005}. When a TF reaches the target it binds with state dependent affinities $\chi_1$ and $\chi_2$. In state 1 a TF switches to state 2 with rate $k_{12}$. In state 2, in addition to switching to state 1 with rate $k_{21}$, a TF can also dissociate with rate $k_{23}$ and switch to state 3 where it diffuses in the bulk. From state 3 it associates with the DNA with the rate $k_{32}$ at a random position drawn from a Gaussian distribution with variance $\sigma_3^2$ centered around the previous dissociation point.

Because of the Gaussian attaching distribution, we model state 3 as an 1D diffusion process along the DNA with an effective diffusion constant $D_3=\frac{\sigma_3^2 k_{32}}{2}$ and no target affinity ($\chi_3=0$). Thus, we finally arrive at a framework with switchings between three 1D states. The backward Fokker-Planck equation for the probability $p(x,t,n|y,m)$ to find the TF at time $t$ in state $n$ at position $x$, conditioned that it started at $t=0$ in state $m$ at position $y$, is \cite{ReingruberHolcman_JCondM2010,ReingruberHolcman_PRL2009}
\bea \label{backwardFPEq_text}
\p_t p(x,t,n|y,m) &=& D_{m} \p^2_{y}  p(x,t,n|y,m) -2\chi_m p(x,t,n|y,m) \delta(y) \nn\\
&& -  \sum_{i=1}^3 k_{mi} \( p(x,t,n|y,m) - p(x,t,n|y,i)\)\,,
\eea
with reflecting boundary conditions at $y=\pm L$. The mean sojourn time spent in state $n$ is
\bea
\tau_{n,m}(y)= \int_0^\infty dt \int_{-L}^L dx \, p(x,t,n|y,m)\,.
\eea
From Eq.~\ref{backwardFPEq_text} we find that the $\tau_{n,m}(y)$ satisfy the system of equations
\bea\label{eqSojournTimes_text}
D_{m}  \tau_{n,m}''(y)- \sum_{i=1}^3  \(k_{m+}\delta_{mi}- k_{mi}\) \tau_{n,i}(y) - 2\chi_m \delta(y) \tau_{n,m}(y)   = -\delta_{nm}
\eea
with $k_{m+} = \sum_{j=1}^3 k_{mj}$. In the Supplementary Information (SI) we exactly solve Eq.~\ref{eqSojournTimes_text} and derive analytic expressions for the $\tau_{n,m}(y)$. The MFPT when initially in state $m$ at position $y$ is $\tau_{m}(y) = \sum_{n=1}^3 \tau_{n,m}(y)$. We focus on the MFPT with uniform initial distribution, $\bar \tau_{m} = \frac{1}{2L}\int_{-L}^L \tau_{m}(y) dy$. Because switchings between states occur fast compared to the overall search time, the dependency of $\bar \tau_{m}$ on $m$ is negligible. Furthermore, the mean sojourn times $\bar \tau_{i,m}$ approximately satisfy the scaling relations (see SI Eq.~40) $\bar \tau_{1,m}:\bar \tau_{2,m}:\bar \tau_{3,m} = 1:\frac{k_{21}}{k_{12}}:\frac{k_{32}}{ k_{23}}$. Hence, the MFPT with uniform initial distribution is well approximated by
\bea\label{mfptGeneralExpr}
\bar \tau \approx  \bar \tau_{1,1}\( 1+ \frac{k_{12}}{k_{21}} + \frac{k_{12} k_{23}}{k_{21}k_{32}} \)
=  N_{12} \( \frac{1}{k_{12}}+ \frac{1}{k_{21}} + \frac{k_{23}}{k_{21}} \frac{1}{k_{32}} \)
\eea
where $N_{12} = \bar \tau_{1,1} k_{12}$ is the average number of switchings between states 1 and 2.

To reveal the scaling of the MFPT as a function of the DNA length $L$, we introduce the scaled length $\hat L = \frac{L}{L_0}$ defined with the reference length $L_0=1bp$. To simplify the discussion, we focus on a scenario with maximal affinity in state 1, $\chi_1=\infty$, in which case case the search is over when the TF encounters the target in state 1 (the analysis in the SI is performed with a general $\chi_1$). For a long DNA ($\hat L^2 \mu_i \gg 1$) we compute in the SI
\bea\label{expressionForN12_text}
N_{12} =  \hat L l_{12} \(\frac{ c_{1,1} }{\mu_1 \sqrt{\mu_1} }
+  \frac{ c_{1,2} }{\mu_2 \sqrt{\mu_2}}\) + \hat L^2 \frac{l_{12} l_{21}l_{32}}{3\beta}
\eea
with the following dimensionless parameters:
\bea
l_{ij} = \frac{L_0^2 k_{ij}}{D_i}\,,  \quad c_{1,1} = l_{21} l_{32} \( \frac{a_1 d}{e} + b_1\)  \,, \quad  c_{1,2} = l_{21}l_{32} \( \frac{a_2 d}{e} + b_2\) \nn \\
\alpha = l_{12} +l_{21}+ l_{23}+l_{32}\,, \quad  \beta = l_{21}  l_{32} + l_{12} ( l_{23}  + l_{32})\,, \quad
\mu_{1/2}= \frac{1}{2}\( \alpha \pm \sqrt{\alpha^2 - 4 \beta } \) \nn\\
 a_1= \frac{ \mu_1 ( l_{12} + l_{21}  -\mu_2 ) }{l_{12}(\mu_1-\mu_2)} ,\,\,
\ds a_2= \frac{ \mu_2 ( l_{12} + l_{21}  -\mu_1 ) }{l_{12}(\mu_2-\mu_1)} , \quad b_1 =  \frac{ l_{32}-\mu_1 }{l_{32}(\mu_1-\mu_2)}, \,\, b_2 =  \frac{ l_{32}-\mu_2 }{l_{32}(\mu_2-\mu_1)} \nn \\
 \kappa_2 =\frac{L_0 \chi_2}{D_2}\,,  \quad d = \ds \frac{1}{l_{32}\kappa_2} -  \( \frac{b_1}{\sqrt{\mu_1}} + \frac{b_2}{\sqrt{\mu_2}}\)\,,\quad
\ds  e = \ds  \frac{a_1}{\sqrt{\mu_1}} +  \frac{a_2}{\sqrt{\mu_2}}  + \frac{l_{21}}{l_{12} \kappa_2}\,.\nn
\eea

\subsection*{Search with uniform redistribution in state 3 and no target affinity in state 2}
With uniform redistributions ($\sigma_3=\infty$) we have $l_{32}=2 L_0^2/\sigma_3^2 =0$. With $\kappa_2=0$ and $l_{32}=0$, Eq.~\ref{expressionForN12_text} simplifies to
\bea\label{N12UnifRedist_text}
N_{12} = \hat L l_{12}\( \frac{l_{12}-\mu_2}{\mu_1-\mu_2}  \frac{1}{\sqrt{\mu_1}}  + \frac{l_{12}-\mu_1}{\mu_2-\mu_1} \frac{1}{\sqrt{\mu_2}} \) \,,
\eea
in agreement with \cite{ReingruberHolcman_PRE2011}. We note that the term $\sim L^2$ in Eq.~\ref{expressionForN12_text} vanished and we now have $N_{12}\sim L$, which leads to a faster search for large $L$. As stated before, Eq.~\ref{N12UnifRedist_text} is valid for $\hat L^2 \mu_i \gg 1$ ($i=1,2$) and therefore cannot be applied for $k_{23}\to 0$. For $k_{23}\to 0$ one eigenvalue, say $\mu_1$, vanishes and the condition $\hat L^2 \mu_1 \gg 1$ is violated. For example, with a fixed $L$ and  $k_{23}\to 0$ the TF remains bound to the DNA. In this case we expect that the MFPT scales $\sim L^2$ and not $\sim L$, which is indeed the case, as can be shown by a refined analysis. However, for any fixed value $k_{23}>0$, by increasing $L$, $N_{12}$ eventually scales $\sim L$ due to the uniform redistributions.

\subsubsection*{Optimal switching scenario in state 2}
When the properties of state 1 and state 3 are fixed (and $D_2$ is fixed), we compute the optimal switching rates $k_{21}$ and $k_{23}$ that minimize the MFPT. We introduce the  parameters
\bea
\sigma_1^2 = \frac{2D_1}{k_{12}}\,, \quad \sigma_2^2 = \frac{2D_2}{k_{21} +k_{23}} \,, \quad  q=\frac{k_{23}}{k_{21} +k_{23}}\,,\quad  \zeta=\frac{\sigma_{1}^2}{\sigma_{2}^2}\,.
\eea
$\sigma_1^2$ is the mean square displacement in state 1, $\sigma_2^2$ is the mean square displacement in state 2 before switching either to state 1 or 2, $q$ is the detaching probability, and $\zeta$ is the ratio of the displacements in state 1 and 2. We use the variables $q$ and $\zeta$ instead of $k_{21}$ and $k_{23}$. We have $l_{23}= l_{12} q\zeta$ and $l_{21}= l_{12} (1-q)\zeta$. Because diffusion in state 1 is slow compared to state 2, we have $\zeta \ll 1$. We further consider that the probability to switch from state 2 to 1 is much larger than the dissociation probability, such that $q\ll 1$. For $\zeta \ll 1$ and $q\ll 1$ we obtain from Eq.~\ref{mfptGeneralExpr} and Eq.~\ref{N12UnifRedist_text} the asymptotic
\bea\label{mfptApprox_3}
\bar \tau  \approx \sqrt 2\frac{L}{\sigma_1}\(  1+ \sqrt{\frac{\zeta}{q}} \)\( \frac{1}{k_{12}}+ \frac{\sigma_1^2}{2D_2 \zeta } + \frac{q}{k_{32}} \)\,.
\eea
For fixed $\sigma_1$, $D_2$, $k_{12}$ and $k_{32}$, the minimum of $\bar \tau$ with respect to $(q,\zeta)$ is
\bea\label{tauOptimalSearch}
\bar \tau =  \frac{ \sqrt 2 L}{\sigma_1 k_{12}}  \( 1+  \gamma \zeta \)^2
\eea
achieved at $\gamma \zeta = \sqrt{\frac{2}{\delta}}$ and $\gamma^2 q\zeta = 1$, where $\gamma = \sqrt{\frac{2 D_2}{k_{32} \sigma_1^2}}$ and $\delta = \frac{k_{32}}{k_{12}} \gamma =  \frac{k_{32}}{k_{12}} \sqrt{\frac{2 D_2}{k_{32} \sigma_1^2}}$. Interestingly, whereas the optimal rate $k_{21}$ depends on the properties of state 1, we find for $k_{23}$ the optimal value $k_{23}= \frac{D_2}{D_1} q \zeta =  \frac{D_2}{D_1}\frac{1}{\gamma^2}= k_{32}$, independent of the properties of state 1.

\subsection*{Search with two states only}
To derive the MFPT with switchings between two states we set $k_{23}\to 0$. With $\kappa_1=\infty$ and $\kappa_2=0$ (state 2 now corresponds to the bulk state without binding) we find
\bea
\bar \tau =\hat L l_{12} \( \frac{l_{21}}{\mu_2} \frac{\hat L}{3}  + \frac{l_{12}}{\mu_2} \frac{1}{\sqrt{\mu_2}} \) \( \frac{1}{k_{12}} + \frac{1}{k_{21}} \)
\eea
with $\mu_2=l_{12}+l_{21}$. For $k_{12}\to 0$ we recover $\bar \tau = \frac{L^2}{3D_1}$. With uniform redistributions in state 2 we get ($l_{21}=0$)
\bea\label{mfpt2States}
\bar \tau = \sqrt{2} \frac{L}{\sigma_1} \( \frac{1}{k_{12}} + \frac{1}{k_{21}} \)\,.
\eea
Eq.~\ref{mfpt2States} as a function of $k_{12}$ has a minimum for $k_{12} = k_{21}$ (compare with $k_{23}=k_{32}$ obtained with 3 states).

\section*{Results}
We present results that we compare to experimental measurements for a dimeric Lac repressor search in E.Coli bacteria. We use a DNA length $2L=4.8 \times 10^6$ bps \cite{Elf_Science2007}, a TF attaches to the DNA from the bulk after an average time $k_{32}^{-1}=1.4 ms$ \cite{MalherbeHolcman_PLA2010,Elf_Science2007}, and the diffusion constant in state 2 is $D_2=2 \mu m^2/s$ ($D_2=1.7 \times 10^7 \frac{bp^2}{s}$) \cite{Elf_Science2007,Hammaretal_Science2012}. We keep these values fixed throughout the following analysis and we focus on investigating the impact of the remaining parameters.

\subsection*{Search scenario with conformational changes}
A TF is freely diffusing in the bulk (state 3) and attaches to the DNA with a Poissonian rate $k_{32}$ (Fig.\ref{fig_scenario}). We consider that the association position follows a Gaussian distribution with variance $\sigma_3^2$ centered around the previous dissociation point. $\sigma_3$ is the correlation distance between subsequent detaching and attaching positions. Hence, $\sigma_3$ is an effective parameters that implicitly depends on the DNA configuration and on coiling. For example, a uniform re-attaching distribution obtained for $\sigma_3=\infty$ is usually attributed to a highly packed DNA conformation \cite{ReingruberHolcman_PRE2011,Benichou_Traps_PRL2009,SheinmanKafri_PhysBiol2009,SlutskyMirny_BJ2004,LomholtMetzler_PRL2005,CoppeyBenichou_TFSearch_BJ2005}. By varying $\sigma_3$ we can investigate how the DNA conformation affects the MFPT. We assume that a TF switches between a stable and an unstable protein conformation. The lifetime of the unstable conformation $\xi^{-1}$ is short such that a TF quickly returns to its stable conformation after a spontaneous conformation change. When a TF is attached to the DNA and in the stable conformation (state 2) it non-specifically interacts with the DNA backbone and diffuses in a smooth potential well with non-specific energy $E_{ns}$ and fast diffusion constant $D_2$. In state 2 a TF can either dissociate from the DNA with rate $k_{23}$, or switch to the unstable conformation (state 1) with rate $k_{21}$. The unstable conformation allows for additional specific TF-DNA interactions that modify the residence time $k_{12}^{-1}$ in state 1. We use the Arrhenius like relation $k_{12}=\xi e^{-\Delta E}$, where $\Delta E=E_{ns}-E$ (in units of $k_B T$). We use a maximal target affinity in state 1, $\chi_1=\infty$, in which case the search is finished when a TF reaches the target in state 1 for the first time. In state 2, the outcome at the target depends on the affinity $\chi_2$ (respectively the dimensionless parameter $\kappa_2$). The search is over for $\kappa_2=\infty$, in which case a TF has maximal target affinity already in state 2. In the opposite case $\kappa_2=0$ a TF has no indication in state 2 that it has reached the target site and there is a probability that he misses the target and detaches without binding. Because switching to state 1 at the target site ends the search, by varying $\kappa_2$ we can explore the impact of induced switchings at the target site.

We use constant switching rates $k_{21}$, $k_{23}$ and $k_{32}$. This is valid for a spatially homogenous DNA and a homogenous non-specific interaction in state 2. In contrast, $k_{12}$ and $\sigma_1$ depend on the specific binding energy $E$ and are therefore not constant along the DNA. To account for this, we first derive results using a constant $E$ corresponding to a homogenous DNA, and in a subsequent step we average using a Gaussian distribution for $E$. Because of strong specific interactions, we focus on displacements $\sigma_1$ that are small. The lower bound for $\sigma_1$ is reached when a TF becomes immobilized in state 1. However, this does not correspond to $\sigma_1=0$, because a TF at least scans the base pair it binds to. A non-zero $\sigma_1$ also accounts for stochastic fluctuations in the DNA position due to the switching process. By noting that the MSD of the maximum displacement of a diffusion process is $2\sigma_1^2$, we use $\sigma_1=\frac{1}{\sqrt{2}}$ to model the limiting case where a TF becomes immobilized in state 1 and scans only a single base pair.

To facilitate the comparison with experimental data, we introduce the following parameters that characterize various properties of a search process:
\bea\label{defAdditionalPArameters}
\tau_{dna}=\frac{1}{k_{23}} + \frac{k_{21}}{k_{23}} \frac{1}{k_{12}} \,, \quad r_{1d3d} =  k_{32}\tau_{dna} \,, \quad \sigma^2_{dna} = \sigma_2^2 \frac{1}{q} + \sigma_{1}^2 \frac{1-q}{q}\,,\quad  D_{dna} = \frac{\sigma^2_{dna}}{2\tau_{dna}}  \,.
\eea
$\tau_{dna}$ is the average time a TF stays bound to the DNA before detaching; $r_{1d3d}$ is the ratio of the time bound to the DNA to diffusing in the cytoplasm; $\sigma^2_{dna}$ is the mean square displacement along the DNA before detaching; $D_{dna}$ is the effective diffusion constant for sliding.

\subsection*{Search with uniform redistributions and no target affinity in state 2}
\begin{figure}[!htb]
\begin{center}
\includegraphics[scale=0.45]{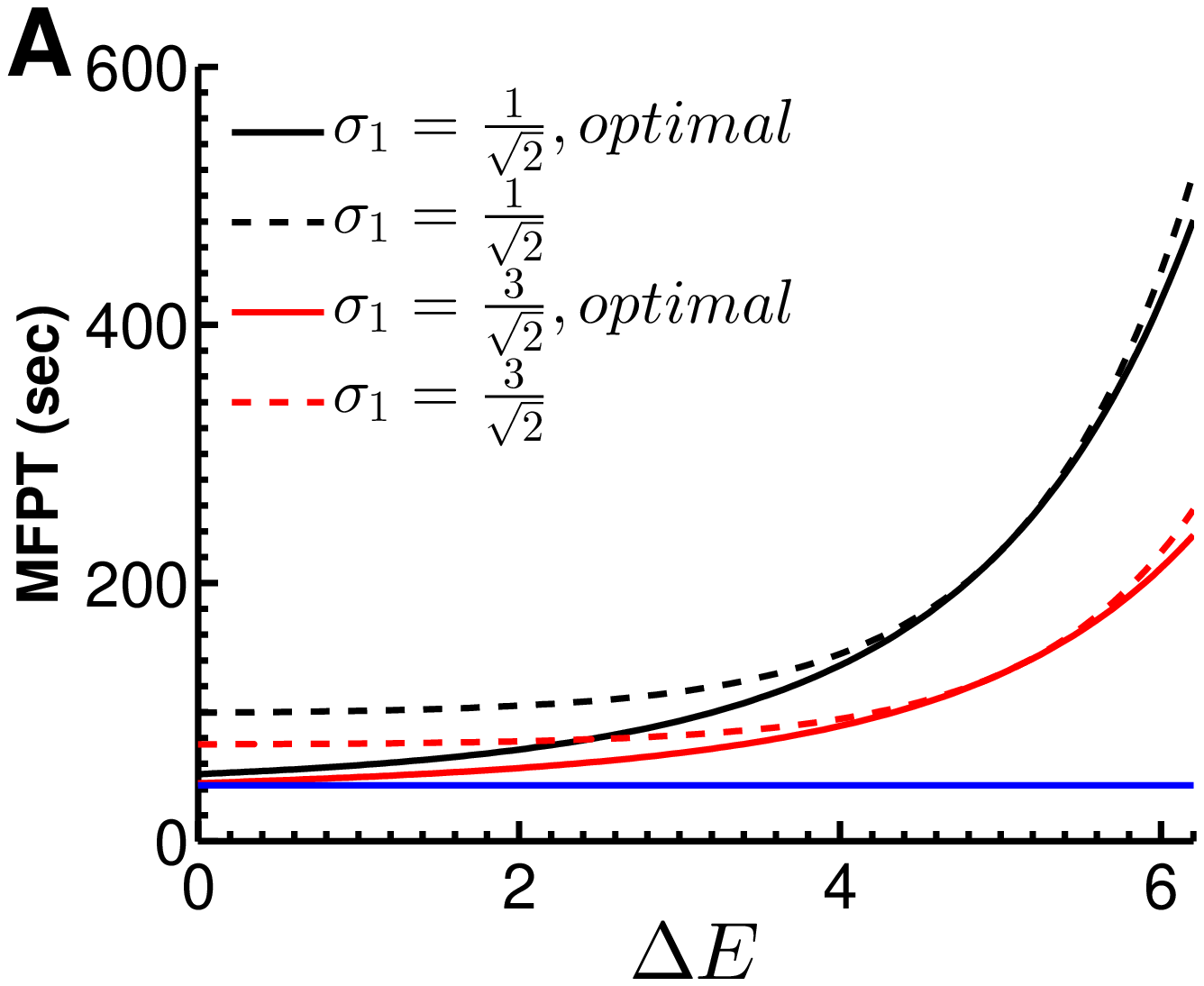}
\includegraphics[scale=0.45]{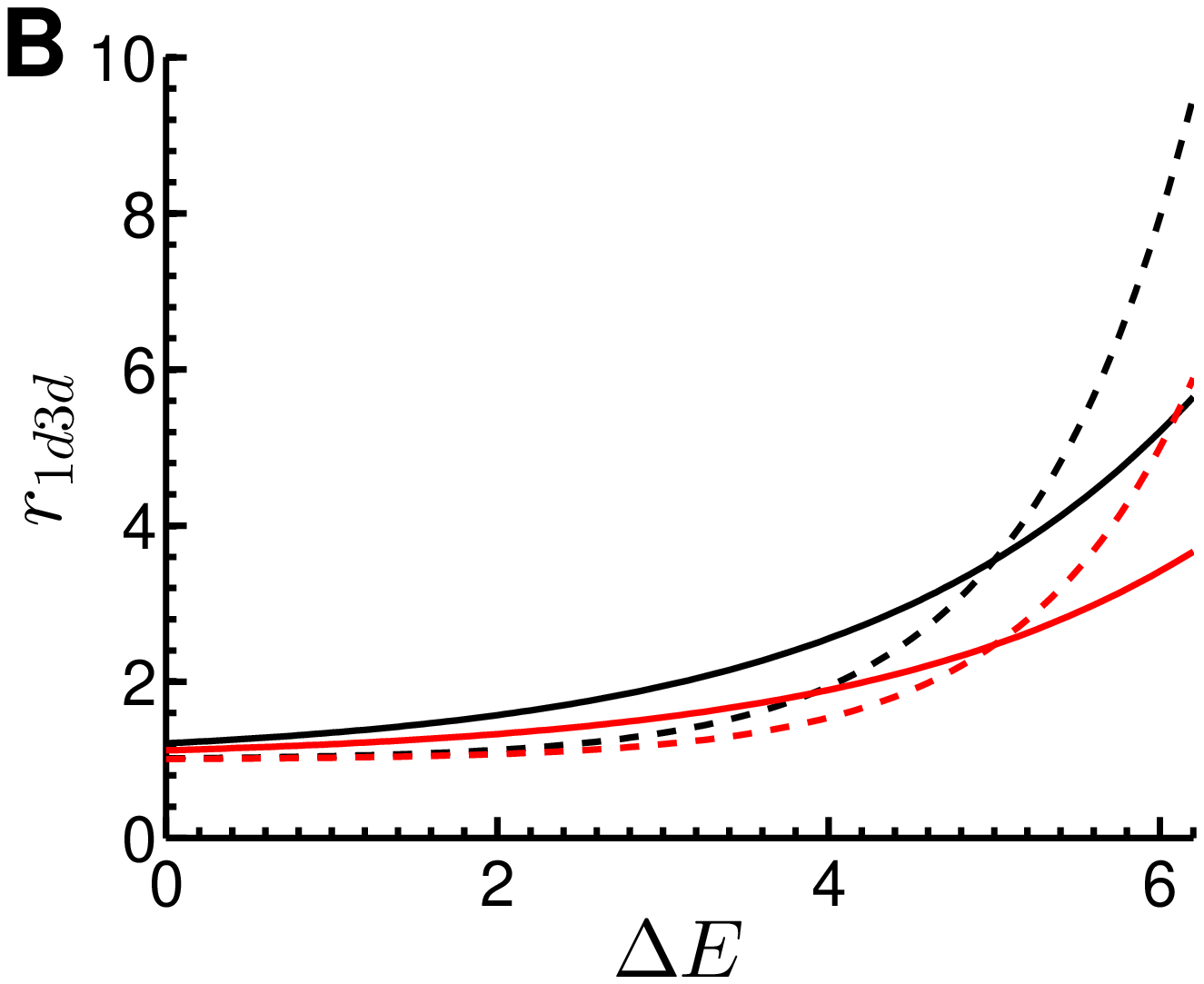}
\includegraphics[scale=0.45]{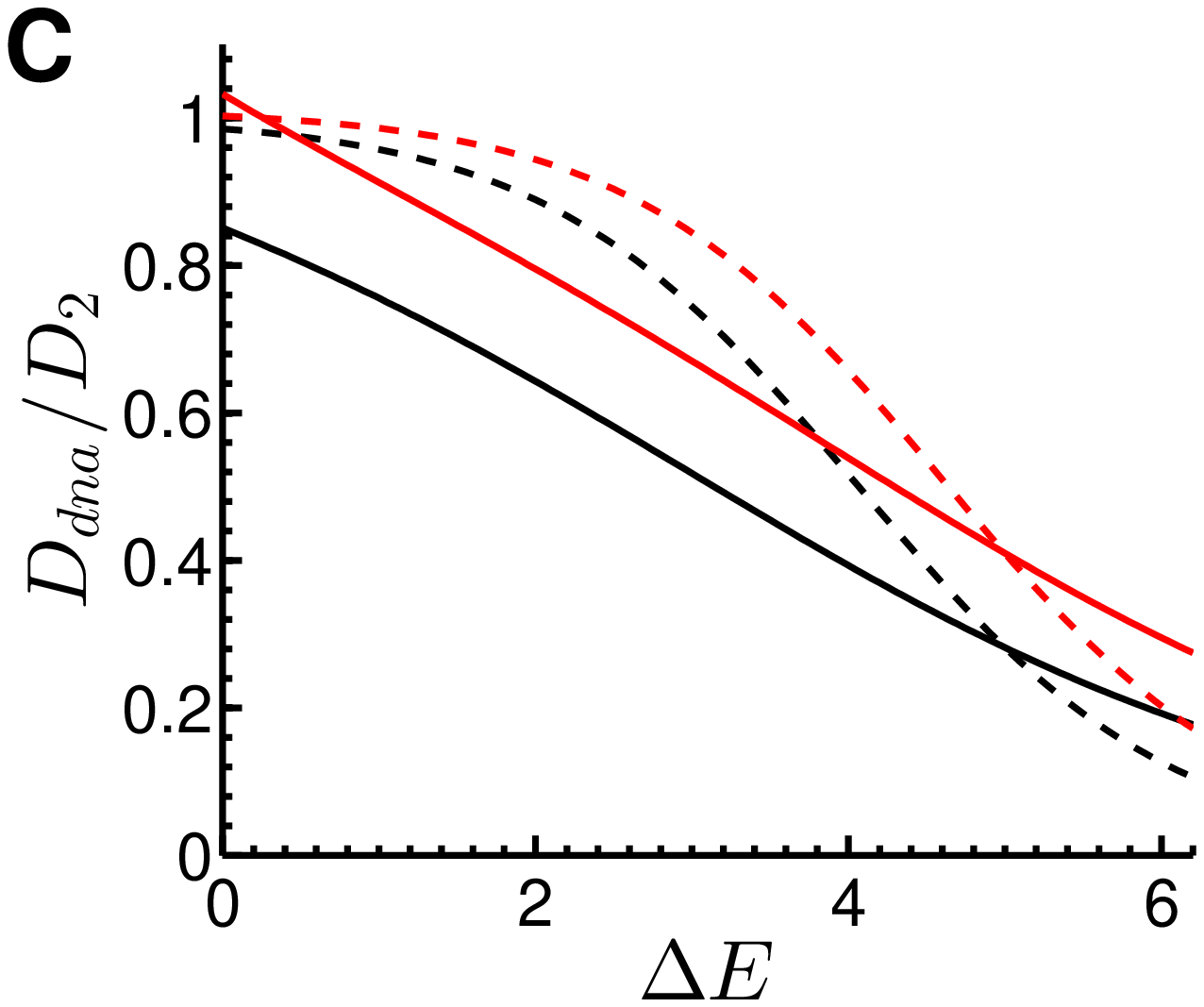}
\includegraphics[scale=0.45]{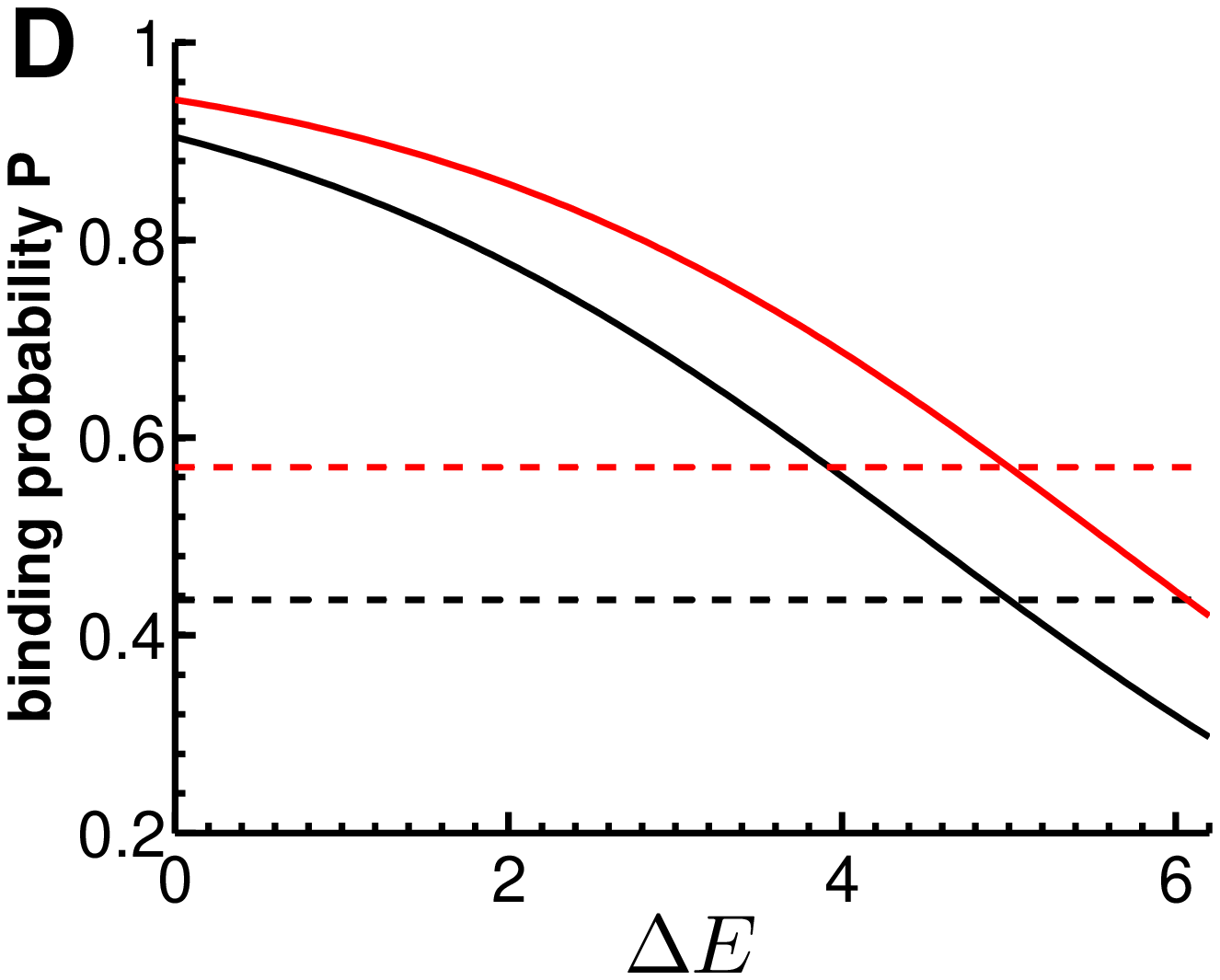}
      \caption{{\bf Search with no target affinity in state 2 ($\kappa_2=0$) and uniform redistributions ($\sigma_3=\infty$).} (A) MFPT for optimal and non-optimal searches as a function of the specific binding strength  for two values of $\sigma_1$ ($k_{12}=\xi e^{-\Delta E}$, $\xi =10^7 s^{-1}$).
      For $\sigma_1=1/\sqrt 2$ a TF is immobile in state 1 and scans only a single bp, for $\sigma_1=3/\sqrt 2$ it scans three bps. The optimal MFPT is computed with Eq.~\ref{N12UnifRedist_text}, the non-optimal with Eq.~\ref{expressionForN12_text}. The blue line corresponds to the MFPT for a two-states switching process between state 2 and 3 with an absorbing target in state 2 ($\kappa_2=\infty$). Energies are in units of $k_B T$. The rest of the parameters are: $\kappa_1=\infty$, $k_{32}^{-1}=1.4ms$, $D_2=2\mu m^2/s$, $k_{23}=k_{32}$. The value of $k_{21}$ for the non-optimal search equals the optimal value computed with $\Delta_E=5$. (B) Ratio of the time spent associated with the DNA compared to freely diffusing in the bulk. (C) Apparent diffusion constant for sliding along the DNA. (D) Probability $P$ to bind to the target before dissociation when arriving at the target site. }
       \label{fig2}
\end{center}
\end{figure}
We start by analyzing search processes as a function of the switching rate $k_{12}$ with no target affinity in state 2 ($\kappa_2=0$) and uniform redistributions in state 3 (Fig.~\ref{fig2}). We write $k_{12}$ as function of the binding strength, $k_{12}=\xi e^{-\Delta E}$, and plot quantities as a function of $\Delta E$. We use the basal rate $\xi = 10^{7}s^{-1}$, which is similar to the attempt frequency $10^8 s^{-1}$ used in \cite{SlutskyMirny_BJ2004}, or $10^{6} s^{-1}$ from \cite{Benichou_Traps_PRL2009}. At this stage the exact value of $\xi$ is not important to show the behaviour as a function of $\Delta E$. For example, a smaller value for $\xi$ would shift the origin of the $\Delta E$-axis to the right in Fig.~\ref{fig2} and Fig.~\ref{fig3}, but otherwise does not affect the graphs. Later on we will estimate a more appropriate value for $\xi$ by considering a Gaussian binding energy distribution.

We compare optimal and non-optimal searches for $\sigma_1=\frac{1}{\sqrt 2}$ (1 bp is scanned in state 1) and $\sigma_1=\frac{3}{\sqrt 2}$ (up to 3 bps are scanned in state 1). The optimal search is characterized by $k_{23}=k_{32}$ and a rate $k_{21}$ that is a function of $\Delta E$ and $\sigma_1$ (see Eq.~\ref{tauOptimalSearch}). For non-optimal searches we use $k_{23}=k_{32}$ and a rate $k_{21}$ that is independent of $\Delta E$, since the properties of state 1 should not affect the switching rate $k_{21}$ in state 2. We use the optimal value for $k_{21}$ computed with $\Delta E=5$ (hence, $k_{21}=1.72\times 10^5s^{-1}$ for $\sigma_1=\frac{1}{\sqrt 2}$ and $k_{21}=9.92 \times 10^4s^{-1}$ for $\sigma_1=\frac{3}{\sqrt 2}$), which gives a fast search also with large $\Delta E$. We use two different values for $\sigma_1=\frac{1}{\sqrt 2}$ and $\sigma_1=\frac{3}{\sqrt 2}$ to facilitate the comparison between optimal and non-optimal curves: in this case the optimal and non-optimal MFPT coincide for $\Delta E=5$ (Fig.~\ref{fig2}A).

Fig.~\ref{fig2}A shows that even an immobilized TF in state 1 can have a MFPT that is compatible with the experimental finding $\sim 350s$ \cite{Elf_Science2007,Hammaretal_Science2012}. The MFPT is faster for $\sigma_1=\frac{3}{\sqrt 2}$ because more DNA is scanned during the same residence time in state 1. Interestingly, the MFPT varies only very little as a function of $\Delta E$ up to values $\Delta E\sim5$ (Fig.~\ref{fig2}A), suggesting that the search is insensitive to a large part of binding energy fluctuations. The asymptotic value of the optimal MFPT for small $\Delta E$ corresponds to a two states process where a TF switches between state 2 and 3 and has maximal target affinity in state 2 (Fig.~\ref{fig2}A, blue curve). This is consistent with results from \cite{ReingruberHolcman_PRL2009} showing that a switching process can have a fast MFPT even if the searcher can only bind in the slow state.

For an optimal search process with only two states (bulk and one sliding state), a TF spends an equal amount of time in the bulk and associated to the DNA. This is not any more the case for an optimal three states process (Fig.~\ref{fig2}B). Although we have $k_{23}=k_{32}$, which is similar to the condition for a two states process, because of switchings to state 1, a TF spends much more time bound to the DNA. For an optimal search we have $r_{1d3d} = 1 + \sqrt{2 \delta}$, where $\delta$ is defined after Eq.\ref{tauOptimalSearch}. For $\Delta E =6$ and $\sigma_1=\frac{1}{\sqrt 2}$ we obtain $r_{1d3d}\approx10$, which is similar to in vivo findings that a dimeric Lac repressor spends ~90\% of the search time bound to the DNA \cite{Elf_Science2007}. The effective diffusion constant for sliding $D_{dna}$ decreases as specific binding becomes stronger (Fig.~\ref{fig2}C). For an optimal search we compute $D_{dna} \approx {D_2}/{r_{1d3d}}$. Thus, for $r_{1d3d}=10$ we obtain $D_{dna} \approx 0.2\frac{\mu m^2}{s}$, which is around 4 times larger than values estimated from single molecule tracking experiments on flow stretched DNA \cite{Elf_Science2007}. However, such a value is in good agreement with results from molecular dynamics simulations for a Lac dimer \cite{Marklundetal_PNAS2013}, and with an apparent 1D diffusion constant $D_{eff} \sim 0.4 \frac{\mu m^2}{s}$ estimated in \cite{Elf_Science2007}. Moreover, a large range of variability is observed for the 1D diffusion constant of a Lac repressor on elongated DNA estimated from single molecule imaging techniques \cite{WangAustinCox_PRL2006}. Whereas $D_{dna}$ strongly depends on $k_{12}$, the sliding distance $\sigma_{dna}$ is not affected by the residence time in state 1 (if $\sigma_1$ remains unchanged) and is determined by diffusion in state 2 and the detaching rate $k_{23}$. For an optimal search with $k_{23}=k_{32}$, by neglecting $\sigma_1$, we obtain $\sigma_{dna}\approx \sqrt{\frac{2D_2}{k_{23}}}\approx 220bp$. This value is much larger than in vivo measurements for a Lac dimer around 40 bps \cite{MahmutovicEtal_NuclAcidRes2015,Hammaretal_Science2012}, but compatible with a value around 240 bps obtained from molecular dynamics simulations \cite{Marklundetal_PNAS2013}. Similar to $D_{dna}$, also for $\sigma_{dna}$ a large experimental variability is observed using single molecule imaging techniques \cite{WangAustinCox_PRL2006}.

Finally, when arriving at the target site in state 2, a TF can as well detach without binding to the target \cite{MahmutovicEtal_NuclAcidRes2015,Hammaretal_Science2012}. To characterize such events, we compute the conditional probability $P$ to bind before detaching when arriving at the target site in state 2 (see SI Eq.~61). The probability depends on the affinity $\kappa_2$, for example, for $\kappa_2=\infty$ we have $P=1$. For $\kappa_2=0$ the probability is not zero and it depends on the local switching dynamics. In general, $P$ can be expressed as a function of the sliding distances independent of the switching rates. Thus, for constant $\sigma_1$, $P$ is independent of $k_{12}$ or $\Delta E$ (Fig.~\ref{fig2}D). For the optimal search process $P$ depends on $\Delta E$ because $k_{21}$ and therefore $\sigma_2$ vary with $\Delta E$.

\subsection*{Search with finite redistributions and induced switchings at the target site}
\begin{figure}[!htb]
\begin{center}
\includegraphics[scale=0.45]{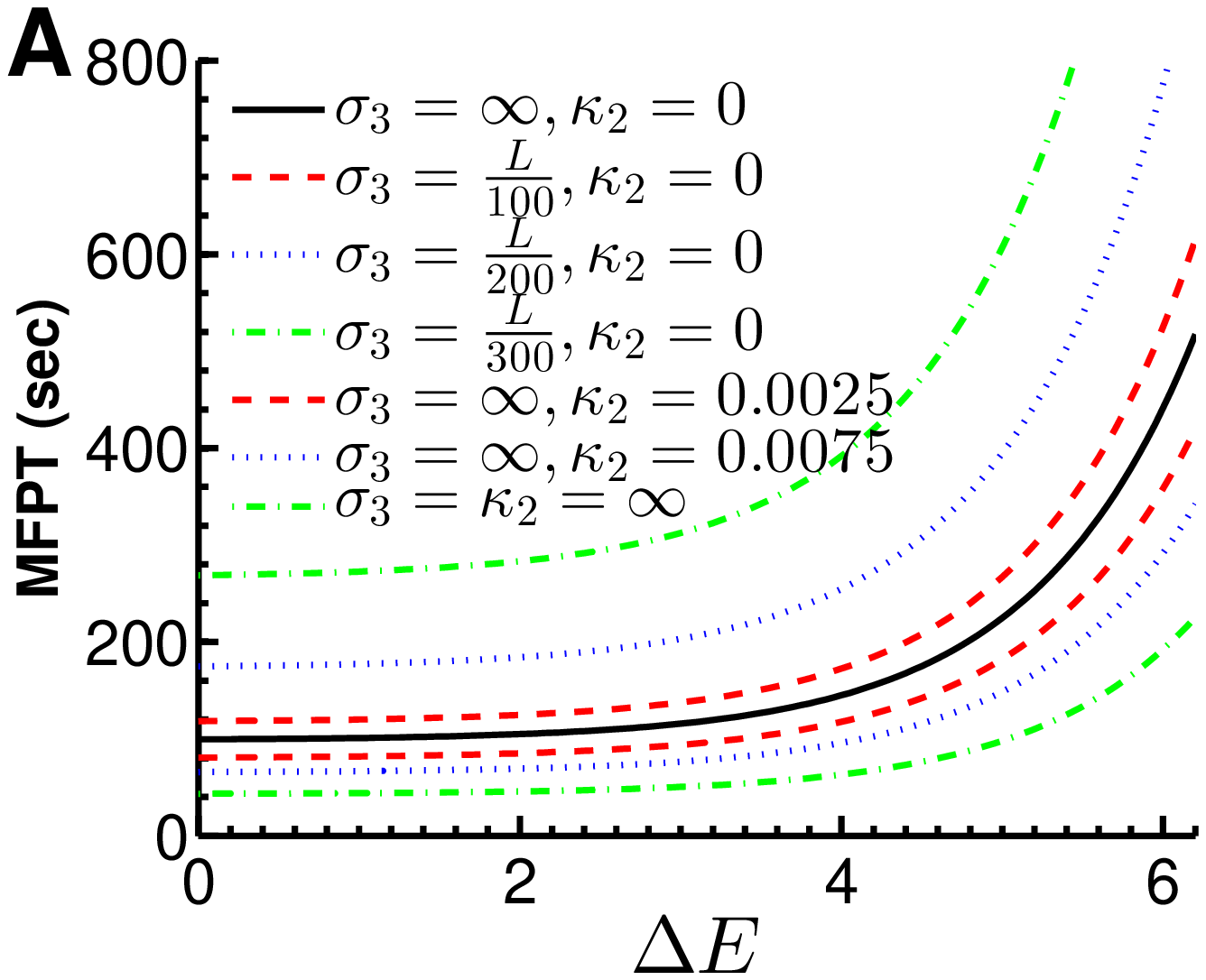}
\includegraphics[scale=0.45]{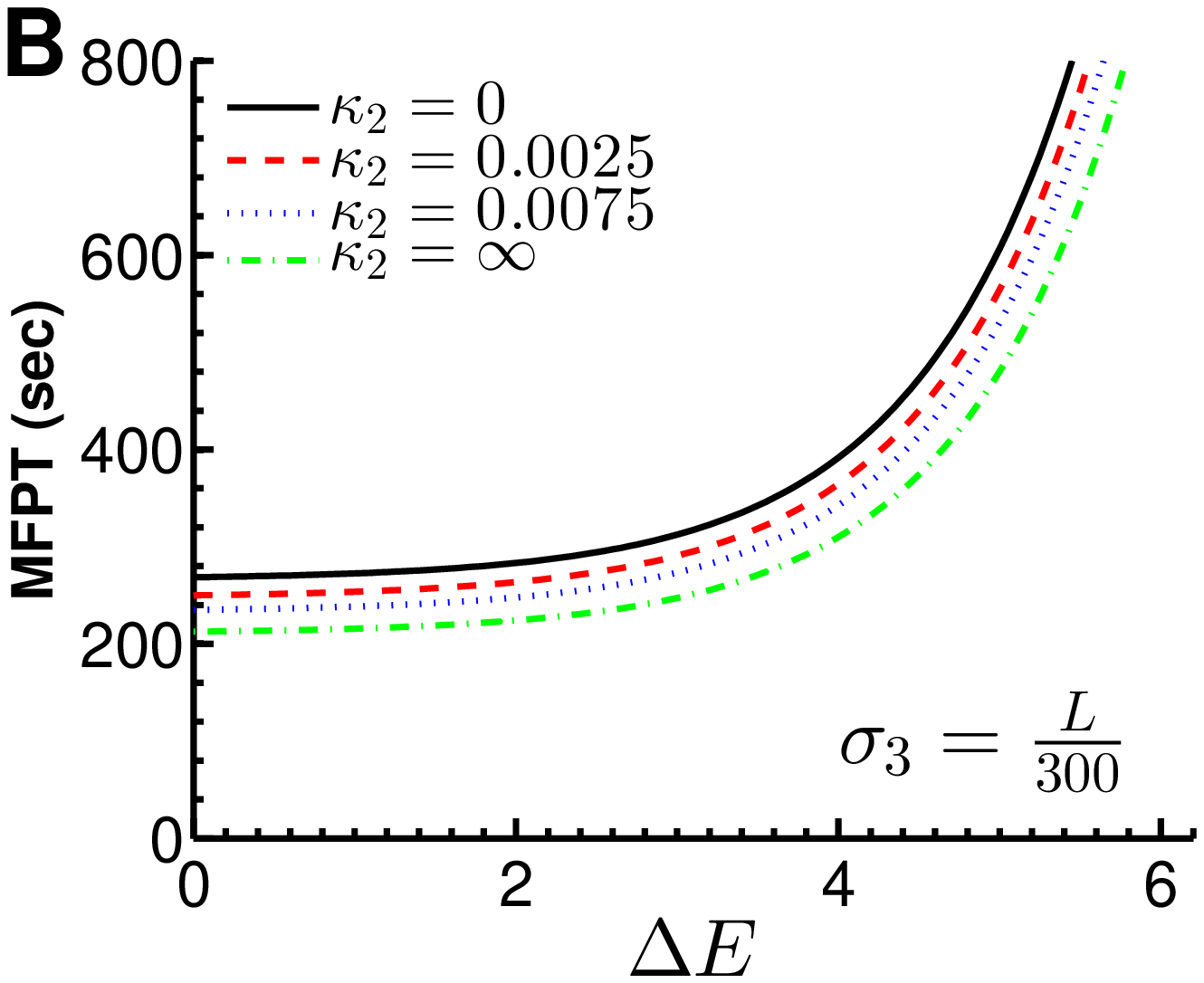}
\includegraphics[scale=0.45]{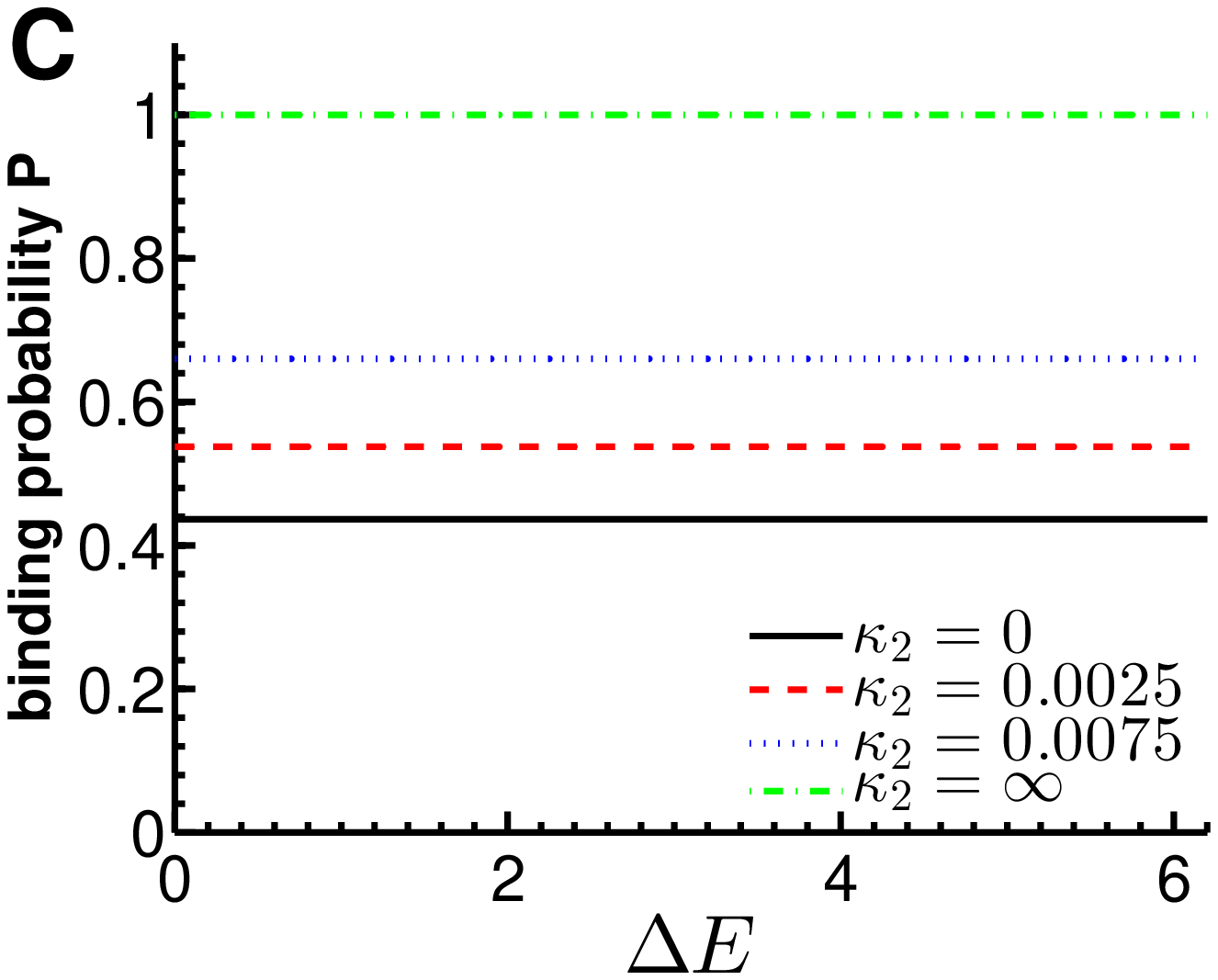}
\includegraphics[scale=0.45]{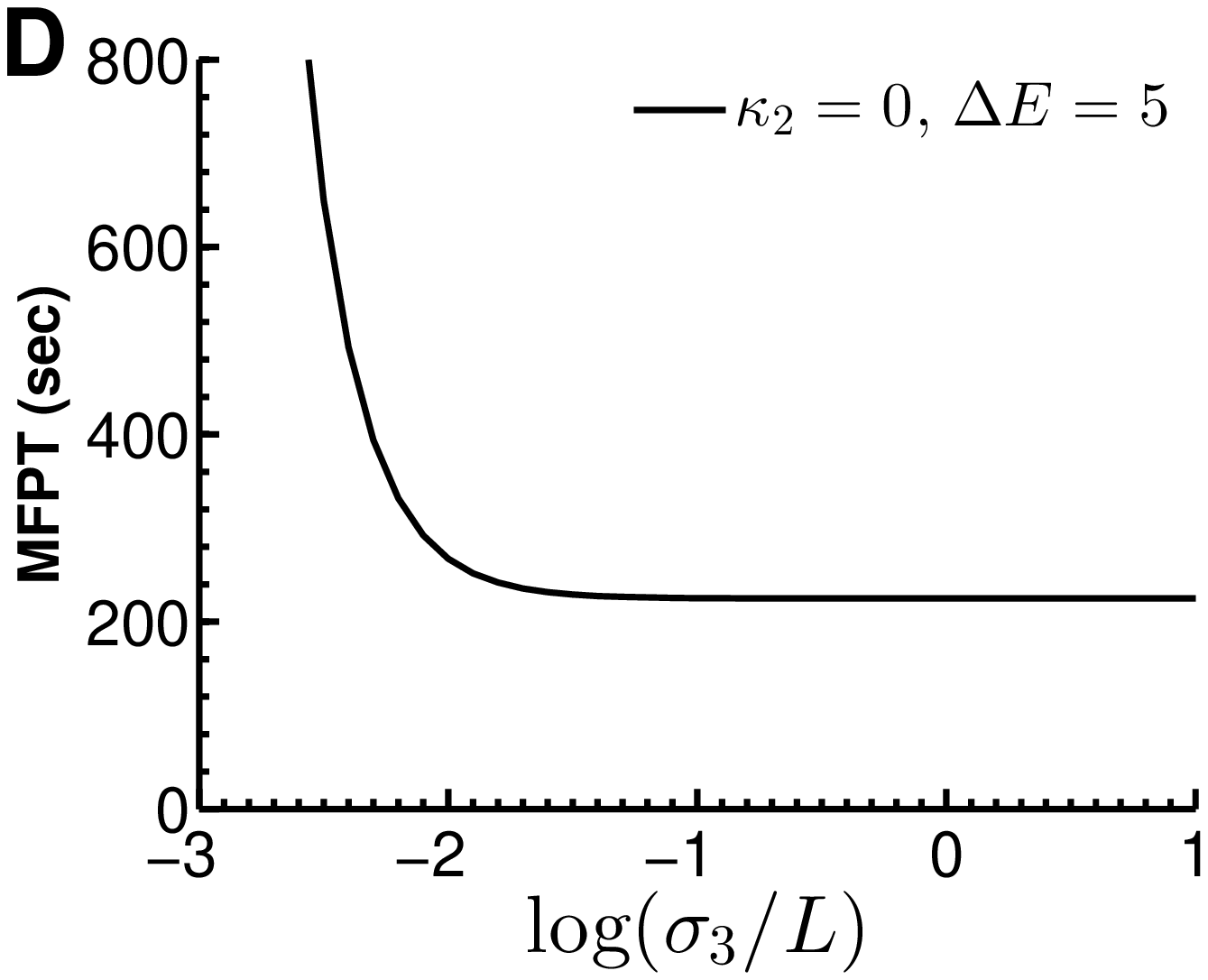}
      \caption{{\bf Search with $\kappa_2>0$ and $\sigma_3<\infty$.} We consider the search from Fig.~\ref{fig2} with $\sigma_1=1/\sqrt 2$ and modify $\kappa_2$ and $\sigma_3$. (A) MFPT for various $\kappa_2$ and $\sigma_3$. (B) MFPT with $\sigma_3=L/300$ and various $\kappa_2$. (C) Probability to bind to the target before dissociation when arriving a the target site for various $\kappa_2$ (note that $P$ is independent of $\sigma_3$). (D) MFPT as a function of $\sigma_3$ for $\kappa_2=0$ and $\Delta E=5$. }
       \label{fig3}
\end{center}
\end{figure}

We proceed and study the effect of the DNA configuration and induced switchings at the target site by varying $\sigma_3$ and $\kappa_2$. We consider a non-optimal search with $\sigma_1=\frac{1}{\sqrt{2}}$. Lowering the correlation distance $\sigma_3$ up to $\sigma_3\sim\frac{L}{100}$ (around 1\% of the genome is correlated) has only little impact on the MFPT (Fig.~\ref{fig3}A,D), but a further decrease strongly increases the MFPT (Fig.~\ref{fig3}D). In contrast, induced switchings at the target site ($\kappa_2>0$) only moderately reduce the MFPT (Fig.~\ref{fig3}A,B). For example, the difference in the MFPT between $\kappa_2=0$ and $\kappa_2=\infty$ is much smaller compared to $\sigma_3\sim\frac{L}{100}$ and $\sigma_3\sim\frac{L}{300}$ (Fig.~\ref{fig3}B). We conclude that a reduced coiling cannot be compensated by induced switchings at the target site. Although a larger $\kappa_2$ increases the probability to bind to the target (Fig.~\ref{fig3}C), this has only a minor impact because $P$ has already a value around 50\% for $\kappa_2=0$.

\subsection*{Search with a Gaussian binding energy distribution}
So far we used a constant rate $k_{12}$ corresponding to a constant binding energy $E$. In reality, $k_{12}$ depends on the DNA sequence and therefore on the DNA position $x$. To account for this, we consider a search with a Gaussian binding energy distribution $\rho(E)$ \cite{SlutskyMirny_BJ2004,Gerlandetal_TFinteraction_PNAS2002}. We further consider the case where a TF is immobile in state 1 such that $\sigma_1=1/\sqrt 2$ and the number of switchings $N_{12}$ (see Eq.~\ref{N12UnifRedist_text}) are both independent of $x$. Let $w(x)$ be the weight function that measures how often position $x$ is visited during a search process compared to the average. For a uniform initial distribution, from symmetry considerations, we can deduce that $w(x)=1$. In this case the MFPT is
\bea
\bar \tau = N_{12} \( \int \frac{w(x)}{k_{12}(x)} dx+ \frac{1}{k_{21}} + \frac{k_{23}}{k_{21}} \frac{1}{k_{32}}\)
=  N_{12} \(  \int \frac{\rho(E)}{k_{12}(E)} dE+ \frac{1}{k_{21}} + \frac{k_{23}}{k_{21}} \frac{1}{k_{32}} \)\,.
\eea
Hence, we can compute the MFPT  with the average switching rate
\bea\label{meanSwitchRate}
\bar k_{12}^{-1} = \int \frac{\rho(E)}{k_{12}(E)} dE = \int \frac{\rho(E)}{\xi e^{E-E_{ns}}} dE\,.
\eea
With $E_{ns}\approx -11$ and a Gaussian distribution $\rho(E)$ with variance $\sigma=5$ \cite{SlutskyMirny_BJ2004,Gerlandetal_TFinteraction_PNAS2002} we obtain $\bar k_{12}\approx \xi e^{-\zeta}$ with $\zeta = \frac{\sigma^2}{2} + E_{ns}\approx 1.5$. Next we checked how the results in Fig.~\ref{fig2} with $\Delta E  = \zeta= 1.5$ comply with the experimental data. We find that a much better level of agreement is achieved for $\Delta E \approx 5.5$. We note that the range of the $\Delta E$ axis in Fig.~\ref{fig2} depends on the value of $\xi$. Had we used a different value $\xi'= \xi e^{-5.5 -\zeta}= \xi e^{-4}\approx 10^5 s^{-1}$, the origin of the $\Delta E$ axis would be shifted to the right and the results for $\Delta E = 5.5$ in Fig.~\ref{fig2} would appear at $\Delta E = \zeta$. Thus, by using $\xi \sim 10^5 s^{-1}$ instead of $\xi \sim 10^7 s^{-1}$ we obtain a search scenario that is compatible with experimental data even in presence of a Gaussian energy distribution with large variance. Moreover, for $\xi \sim 10^5 s^{-1}$ we also have $\xi\sim k_{21}$ and we have the Arrhenius like relation $\frac{k_{12}}{k_{21}}\approx e^{-(E-E_{ns})}$.

\section*{Discussion}
We investigated a framework for facilitated diffusion with switchings between three states: a bulk state (state 3) and two states with sliding along the DNA (state 1 and 2) motivated by two TF protein conformations. The TF-DNA interaction and the target affinity depend on the conformation. From the bulk, a TF associates to the DNA with a Poissonian rate and a Gaussian distribution centered around his previous dissociation point. We analytically computed the MFPT, and the conditional probability to bind to the target before detaching when arriving at the target site. We further defined and computed various other properties that characterize the search process, e.g. sliding length, effective 1D diffusion constant or ratio of the time spent in 1D compared to 3D. We compared our results with experimental data for the dimeric Lac repressor search in E.Coli bacteria. We investigated various properties of a search process that we now discuss in more detail.

\subsection*{Impact of the DNA conformation}
It is still largely unclear how strongly the DNA conformation affects the search time \cite{Koslover_BJ2011,LomholtBroeckMetzler_PNAS2009,LomholtMetzlerKlafter_PNAS2008,HuGrosbergShlovskii_BJ2006}. In the literature one can find analytic results for two opposite cases: a rod-like DNA or a maximally coiled DNA where the re-attaching distribution is uniform. However, a systematic and consistent analysis where the impact of coiling is gradually changed is still outstanding. In our model, the association rate $k_{32}$ and the correlation distance $\sigma_3$ are two effective parameters that implicitly depend on the DNA conformation. For fixed $\sigma_3$, a higher attaching $k_{32}$ decreases the MFPT, but only up to a lower limit that is attained for instantaneous jumps $k_{32}=\infty$ (Eq.~\ref{mfptGeneralExpr}). The MFPT is minimal for a uniform redistribution ($\sigma_3\to \infty$, Fig.~\ref{fig3}A), a scenario that is frequently used to analyze a facilitated diffusion process with a highly packed DNA conformation \cite{ReingruberHolcman_PRE2011,SheinmanKafri_PhysBiol2009,SlutskyMirny_BJ2004,LomholtMetzler_PRL2005,CoppeyBenichou_TFSearch_BJ2005,Benichou_Traps_PRL2009}. We find that around 1\% of the DNA has to become correlated by the 3D excursions in order to maintain such a fast MFPT (Fig.~\ref{fig3}A). At lower correlation distances the search time is greatly prolonged (Fig.~\ref{fig3}D). The value of $\sigma_3$ also determines how the search time scales as a function of the DNA length $L$. To show this we consider the number of switchings $N_{12}$ that are necessary to find the target. $N_{12}$ increases proportional to $L$ for $\sigma_3\to \infty$ (Eq.~\ref{expressionForN12_text}), and such a linear dependency is usually assumed in the literature. However, for finite $\sigma_3$, the leading order asymptotic for large $L$ is $N_{12} \sim L^2$ and not $N_{12}\sim L$. For finite $L$, a careful analysis is needed to determine whether the contribution $\sim L$ or $\sim L^2$ is dominant. For our analysis we considered that $\sigma_3$ and $k_{32}$ are independent parameters, however, in general their values will be correlated. For example, lets consider stretched DNA. In this case, by assuming a correlation distance $\sigma_3\sim L$, we find $N_{12}\sim L$. However, because the DNA is stretched and a TF is diffusing with diffusion constant $D_3$ in the bulk, we must at least have $k_{32}^{-1}\sim \sigma_3^2/D_3$. Finally, this leads to a MFPT that scales $\sim L^2$ and not $\sim L$. On the other hand, with strong coiling one might have $\sigma_3\sim L$ with a fast rate $k_{32}$ that is almost independent of $\sigma_3$, such that the MFPT scales $\sim L$. We conclude that without coiling it is not possible to have a MFPT that scales $\sim L$. Coiling is permissive to obtain at the same time a large correlation distance $\sigma_3$ and a fast attaching rate $k_{32}$.

\subsection*{Impact of induced switchings}
If switchings from state 2 to state 1 are induced at the target site, the spontaneous switching $k_{21}$ can be reduced leading to a faster search because a TF spends more time in the fast state 2 \cite{SlutskyMirny_BJ2004,Zhou_SwitchinTFsearch_PNAS2011}. However, it is unclear which physical mechanism would provide such a specificity. We investigated the impact of induced switchings by varying the target affinity in state 2 ($\kappa_2$). We find that a MFPT compatible with experimental data can be achieved without induced switchings  (Fig.~2\ref{fig2}A and Fig.~2\ref{fig3}A). We estimated that in this case a switching rate around $k_{21}\approx 10^5s^{-1}$ is necessary. This implies conformational changes in the submillisecond range, that have also been suggested in \cite{Tafvizietal_P53SingleMolecule_PNAS2011}. The rate $k_{21}$ could be further reduced by assuming a larger sliding distance $\sigma_1$. If spontaneous switchings to state 1 are fast, the conditional probability $P$ to bind to the target before detaching is already large for $\kappa_2=0$ (Fig.~2\ref{fig3}C). In such a case additional induced switchings ($\kappa_2>0$) do not much affect the search time (Fig.~\ref{fig3}B). Clearly, the impact of induced switchings  would be much larger if $P$ would be small for $\kappa_2=0$. For example, this could be achieved by lowering the target affinity in state 1 ($\kappa_1 < \infty$), or by reducing the switching rate $k_{21}$. With such conditions the MFPT would be strongly increased by blocking induced switchings. In general, for large $\kappa_2=\infty$ the fastest MFPT is achieved by simply not switching to the slow state 1 ($k_{21}=0$). But in this case we return to a two states model with a bulk and a fast sliding state, incompatible with a large binding energy variability.

\subsection*{Search in presence of a Gaussian binding energy profile}
State 1 is characterized by the displacement $\sigma_1$ and the residence time $k_{12}^{-1}$. We analyzed the limiting case where a TF becomes immobilized in state 1 such that it scans only the base pair to which it binds to ($\sigma_1=1/\sqrt 2$). The rate $k_{12}=\xi e^{E-E_{ns}}$ depends on the specific energy $E$ in state 1 and the non-specific energy $E_{ns}$ in state 2. With $\xi = 10^5s^{-1}$, $E_{ns}=-11k_BT$ and a Gaussian distribution for $E$ with variance $\sigma=5k_BT$ we obtained a MFPT around 5-6 minutes, compatible with in vivo experimental data for the dimeric Lac repressor \cite{Elf_Science2007}. It is found that the Lac repressor dimer stays bound to the promoter for an average time $\tau_{b}$ around 5 minutes \cite{Hammaretal_NatGen2014}. In our model this would correspond to a target energy $E=E_{ns}-\ln(\xi \tau_{b}) \approx -28 k_BT$, compatible with data  \cite{SlutskyMirny_BJ2004,Gerlandetal_TFinteraction_PNAS2002}. At strong noncognate DNA sites with $E \sim -25k_BT$, a TF would be trapped only for a short time $k_{12}^{-1}= (\xi e^{-25+11})^{-1}\sim 12s$, which resolves the trapping problem \cite{Benichou_Traps_PRL2009}. The speed-stability paradox strongly relies on the assumption that a TF is found with high probability bound to the target at thermodynamic equilibrium. However, when the MFPT is of the order of minutes, such a high probability implies that a TF blocks the promoter for a very long time. This would impede a fast cellular response, and generate the opposite problem of how a promoter can get rid of a tightly bound TF.

\subsection*{Conclusion and prospects}
In this work we presented a MFPT analysis for a facilitated diffusion search process with switchings between three states: a bulk state and two sliding states where the TF is attached to the DNA. The model is microscopically motivated and describes the local dynamics using effective parameters. Parameter values have to be extracted from more detailed models of the TF-DNA interaction, or by fitting our analytic expressions to experimental data for the dimeric Lac repressor. We focused on a qualitative analysis of the model and we showed that the model predictions account for many features that are observed experimentally.

A major simplification of the current model is the fact that we reduce the impact of the 3D dynamics to a Poissonian association rate $k_{32}$ and a Gaussian re-attaching distribution with width $\sigma_3$. However, this simplification allowed to derive analytic results, which are important to precisely analyze the parameter space. Furthermore, we generalized results with uniform redistributions corresponding to $\sigma_3=\infty$. Assuming that $\sigma_3$ is correlated to the amount of DNA coiling, we could systematically investigate the impact of the DNA conformation. However, polymer models show that the distribution of $\sigma_3$ is not a Gaussian but decays like a power law at large distances
\cite{FuldenbergMirny_ReviewChromatinStructure_2012,Diazdelarosa_BJ2010,Mirnyetal_JPhysA2009}. The re-entry distribution is also more complicated than a single exponential. It will be interesting to investigate in future work how more accurate assumptions for the 3D dynamics based on polymer models change the results presented here. Another interesting project is to compute the MFPT with L\'evy flights in state 3 \cite{LomholtMetzlerKlafter_PNAS2008,LomholtMetzler_PRL2005}.

Instead of using a single state for the 3D dynamics with complex distributions for attaching time and position, one could break down the 3D dynamics into several states with simpler distributions and enlarge the current model by additional states. Each state would account for different properties of the search process, e.g. hoppings, jumps, intersegment and intersegmental transfers.

\section*{Author Contributions}
J.R. conceived and supervised the research; J.C. and J.R. performed the analysis; J.R. and J.C. wrote the paper.

\section*{Acknowledgements}
J. C. acknowledges support from a PhD grant from the University Pierre et Marie Curie.


\cleardoublepage

\section*{Supplementary Information}

\section*{Derivation of sojourn times and MFPT}
We start from the equations for the sojourn times
\bea\label{eqSojournTimes}
D_{m}  \tau_{n,m}''(y)
 - \sum_{i=1}^3   \mbox{K}_{mi} \tau_{n,i}(y) - 2\chi_m \delta(y-y_0) \tau_{n,m}(y) = -\delta_{nm}
\eea
with the switching matrix ($k_{m+} = \sum_{j=1}^3 k_{mj}$)
\bea
K_{mi}  = k_{m+}\delta_{mi}- k_{mi} =
\begin{pmatrix}
k_{12} & -k_{12} &0 \\
-k_{21} & k_{21}+k_{23} & -k_{23} \\
0 & -k_{32} & k_{32} \label{switchMatrix}\\
\end{pmatrix}
\eea
and reflecting boundary conditions at $y=\pm L$. Because the target is located in the center at $y_0=0$ we can restrict the analysis to the region $0\le y\le L$. By integrating Eq.~\ref{eqSojournTimes} around $y=0$ we obtain ($\tau_{n,m}(y)=\tau_{n,m}(-y)$)
\bea\label{boundCondSojTimes}
D_m \tau_{n,m}'(y)|_{y=0^+} = \chi_m \tau_{n,m}(0)
\eea
which are partially reflecting boundary conditions. We remove the killing term in Eq.~\ref{eqSojournTimes} and replace it with these partially reflecting boundary conditions. We introduce the dimensionless position $x=\frac{y}{L}$, the diffusion rates $\nu_m=\frac{D_m}{L^2}$, the dimensionless parameters $\kappa_m=\frac{L \chi_m}{D_m}$, the scaled switching rates $l_{mi}= \frac{k_{mi}}{\nu_m}=\frac{L^2 k_{mi}}{D_m}$ and the scaled switching matrix $L_{mi} = K_{mi}/\nu_m$. The scaled sojourn times
\bea\label{defScaledSojTime}
\hat \tau_{n,m}(x) = \nu_n \tau_{n,m}(x)
\eea
satisfy the system of equations equations ($0\le x\le 1$)
\bea\label{eqSojournTimes2}
\hat \tau_{n,m}''(x)  - \sum_{i=1}^3 L_{mi} \,\hat\tau_{n,i}(x) = - \delta_{nm}
\eea
with reflecting conditions $\hat\tau_{n,m}'(1)=0$ at $x=1$, and partially reflecting conditions  $\hat\tau_{n,m}'(0) = \kappa_m \hat\tau_{n,m}(0)$ at $x=0$.  In state 3 we have a reflecting boundary condition at $x=0$ and $x=1$. The functions ($\bar {\hat \tau}_{n,m} = \int_0^1 \hat \tau_{n,m}(x) dx$)
\bea
v_{n,m}(x) = \hat \tau_{n,m}(x)- \bar {\hat \tau}_{n,m}
\eea
have zero mean and satisfy the system of equations
\bea\label{eqSojournTimes3}
v_{n,m}''(x)  - \sum_{i} L_{mi} \, v_{n,i}(x) = - v_{n,m}'(0)\,.
\eea
The matrix $\mbox{L}_{mi}$ is singular and one eigenvalue is zero. The left eigenvector to the zero eigenvalue is
\bea\label{defFn}
\vec f =\( \frac{l_{21}}{l_{12}} l_{32} ,  l_{32}, l_{23} \)\,.
\eea
From Eq.~\ref{eqSojournTimes2} we obtain
\bea
\sum_{m=1}^3 f_{m} \hat \tau_{n,m}''(x)  = \sum_{m=1}^3 f_{m} v_{n,m}''(x) =  - f_{n}\,, \nn
\eea
and after integration we find
\bea\label{relationBetweenSojTimes}
\sum_{m=1}^3 f_{m} v_{n,m}(x) =  -f_n g(x) = -g_n(x)\,,
\eea
where
\bea\label{defFuncG}
g(x) = \( \frac{(x-1)^2}{2} - \frac{1}{6}\)\,.
\eea
With Eq.~\ref{relationBetweenSojTimes} we express $v_{n,2}(x) -  v_{n,3}(x)$ as a function of $v_{n,1}(x)$ and $v_{n,2}(x)$
and then obtain closed system of equations for $v_{n,1}(x)$ and $v_{n,2}(x)$. By introducing
\bea
w_n(x) = v_{n,1}(x) - v_{n,2}(x)
\eea
we find from Eq.~\ref{eqSojournTimes3}
\bea\label{eqToSolve}
\begin{pmatrix}
v_{n,1}(x) \\
w_n(x)
\end{pmatrix}'' - M \begin{pmatrix}
v_{n,1}(x) \\
w_n(x
\end{pmatrix} =
-\begin{pmatrix} v_{n,1}'(0) \\
w_n'(0) + g_n(x)
\end{pmatrix}
\eea
with
\bea
\mbox{M} =
\begin{pmatrix}
\frac{}{}0 & l_{12} \\ -\frac{\beta}{l_{12}} & \alpha
\end{pmatrix}
\eea
and
\bea
\alpha = l_{12} +l_{21}+ l_{23}+l_{32} \,,\quad \beta = l_{21}  l_{32} + l_{12} ( l_{23}  + l_{32})  \,.
\eea
We solve these equations as a function of $v_{n,1}'(0)$ and $w_n'(0)$ and then compute $v_{n,1}'(0)$ and $w_n'(0)$ using the boundary conditions. The eigenvalues and eigenvectors M of are
\bea
\mu_{1/2} = \frac{1}{2}\( \alpha \pm \sqrt{\alpha^2 - 4 \beta } \) ,\quad
\vec e_i =
\begin{pmatrix}
 \frac{l_{12}} {\mu_i}\\  1
\end{pmatrix}\,,
\eea
with $\beta = \mu_1\mu_2$ and $\alpha = \mu_1 +\mu_2$. With the expansion
\bea\label{defu1u2}
\begin{pmatrix}
v_{n,1}(x) \\
w_n(x
\end{pmatrix} =
u_1(x) \vec  e_1 + u_2(x) \vec  e_2
\eea
and
\bea
\begin{array}{rcl}
\begin{pmatrix} 1 \\  0 \end{pmatrix}  = \ds \frac{\beta }{l_{12}(\mu_2-\mu_1)} \vec e_1  + \frac{\beta }{l_{12}(\mu_1-\mu_2)} \vec e_2  \,, \quad
\begin{pmatrix} 0 \\  1 \end{pmatrix}  = \ds \frac{\mu_1 }{\mu_1-\mu_2} \vec e_1  + \frac{\mu_2 }{\mu_2-\mu_1} \vec e_2
\end{array} \nn
\eea
we derive from Eq.~\ref{eqToSolve}
\bea
u_1''(x) - \mu_1 u_1(x) = -  \( v_{n,1}'(0) \frac{ \beta}{l_{12}(\mu_2-\mu_1)} +  (w_n'(0)   + g_n(x)) \frac{ \mu_1 }{\mu_1-\mu_2} \) \,.
\eea
The equation for $u_2(x)$ is obtained by interchanging $\mu_1$ and $\mu_2$. The function $\tilde u_1(x) = u_1(x) - \frac{g_n(x) }{\mu_1-\mu_2}$ satisfies $\tilde u_1''(x) - \mu_1 \tilde u_1(x) = -c_{n,1}$ with
\bea\label{defC1}
 c_{n,1}= v_{n,1}'(0) \frac{  \beta}{l_{12}(\mu_2-\mu_1)} + w_n'(0) \frac{ \mu_1 }{\mu_1-\mu_2} +\frac{f_n}{\mu_1-\mu_2}\,.
\eea
With $\int_0^1 u_{1}(x) dx  =\int_0^1 \tilde u_{1}(x) dx =0$ we find
\bea\label{solU1}
u_{1}(x) =  c_{n,1} \( \frac{1}{\mu_1}  - \frac{\cosh(\sqrt{\mu_1} (1-x))}{\sqrt{\mu_1} \sinh\sqrt{\mu_1}} \)  + \frac{g_n(x) }{\mu_1-\mu_2}\,.
\eea
The solution $u_2(x)$ is obtained from Eq.~\ref{solU1} by interchanging $(c_{n,1},\mu_1)$ with $(c_{n,2},\mu_2)$, where  $c_{n,2}$ is defined in Eq.~\ref{defC1} with $\mu_1$ and $\mu_2$ interchanged. From Eq.~\ref{defu1u2} we obtain
\bea
v_{n,1}(x) &=& c_{n,1} \frac{l_{12}}{\mu_1 } \( \frac{1}{\mu_1} - \frac{\cosh(\sqrt{\mu_1} (1-x))}{\sqrt{\mu_1} \sinh\sqrt{\mu_1}}  \) +  c_{n,2} \frac{l_{12}}{\mu_2 } \( \frac{1}{\mu_2} - \frac{\cosh(\sqrt{\mu_2} (1-x))}{\sqrt{\mu_2} \sinh\sqrt{\mu_2}}  \)\nn\\
&&  - \frac{l_{12}f_n}{\beta} g(x) \label{eqforVn1}\\
w_n(x) &=& c_{n,1}  \( \frac{1}{\mu_1} - \frac{\cosh(\sqrt{\mu_1} (1-x))}{\sqrt{\mu_1} \sinh\sqrt{\mu_1}}  \) +  c_{n,2}  \( \frac{1}{\mu_2} - \frac{\cosh(\sqrt{\mu_2} (1-x))}{\sqrt{\mu_2} \sinh\sqrt{\mu_2}}  \)
\label{eqforWn}
\eea
From Eq.~\ref{relationBetweenSojTimes} we further find (with $\sum_n f_n =\frac{\beta}{l_{12}}$)
\bea \label{vn1-vn3}
\begin{array}{rcl}
\ds v_{n,1}(x)- v_{n,3}(x) &=& \ds \frac{\beta}{l_{12}l_{23}}  v_{n,1}(x) - \frac{l_{32}}{l_{23}} w_n(x) +\frac{f_n g(x)}{l_{23}}\\
\ds v_{n,2}(x)- v_{n,3}(x) &=& \ds v_{n,1}(x)- v_{n,3}(x) - w_n(x)
\end{array}
\eea

We complete the analysis by computing the values of $v_{n,1}'(0)$ and $v_{n,2}'(0)$. With $v_{n,3}'(0)=0$ and ($g_n'(0) = -f_n$)
\bea\label{relationBetweenSojTimesPrime}
\sum_{m=1}^3 f_{m} v_{n,m}'(0) = f_n
\eea
we can express $v_{n,2}'(0)$  as a function of $v_{n,1}'(0)$
\bea\label{relationv1v2}
\begin{array}{rcl}
\ds  v_{n,2}'(0) = \ds - \frac{l_{21}}{l_{12}} v_{n,1}'(0)  + \frac{f_n}{l_{32}} \,.
\end{array}
\eea
From Eq.~\ref{defC1} and Eq.~\ref{relationv1v2} we get
\bea\label{eqforCn1Cn2_1}
\begin{array}{rcl}
\ds c_{n,1}  &=& \ds  v_{n,1}'(0) a_1 + f_n b_1 \\ \\
 \ds  c_{n,2} &=& \ds  v_{n,1}'(0) a_2 + f_n b_2
\end{array}
\eea
with
\bea\label{defA1A2B1B2}
\begin{array}{c}
\ds a_1= \frac{ \mu_1 ( l_{12} + l_{21}  -\mu_2 ) }{l_{12}(\mu_1-\mu_2)} ,\quad b_1 =  \frac{ l_{32}-\mu_1 }{l_{32}(\mu_1-\mu_2)} \\
\ds a_2= \frac{ \mu_2 ( l_{12} + l_{21}  -\mu_1 ) }{l_{12}(\mu_2-\mu_1)} ,\quad b_2 =  \frac{ l_{32}-\mu_2 }{l_{32}(\mu_2-\mu_1)}\,.
\end{array}
\eea
To derive an equation for $v_{n,1}'(0)$ we compute $w_n(0)$ using Eq.~\ref{eqforWn} and the relation
\bea\label{tau1MinusTau2}
\bar {\hat \tau}_{n,1} - \bar {\hat \tau}_{n,2} =  - \frac{v_{n,1}'(0)}{l_{12}} + \frac{\delta_{n,1}}{l_{12}}
\eea
obtained by integrating Eq.~\ref{eqSojournTimes2}. We find
\bea\label{wn01}
\begin{array}{rcl}
 \ds w_n(0) &=& \ds \frac{c_{n,1}}{\mu_1} +\frac{c_{n,2}}{\mu_2} - c_{n,1}\xi_1 - c_{n,2}\xi_2
=  \frac{v_{n,1}'(0)}{l_{12}} - \frac{f_n}{\beta} - c_{n,1}\xi_1 - c_{n,2}\xi_2 \\ \\
 \ds w_{n}(0) &=& \ds \hat \tau_{n,1}(0)- \hat \tau_{n,2}(0) - (\bar {\hat \tau}_{n,1} - \bar {\hat \tau}_{n,2} )
=  \frac{v_{n,1}'(0)}{\kappa_1}  - \frac{v_{n,2}'(0)}{\kappa_2}  + \frac{v_{n,1}'(0) }{l_{12}} - \frac{\delta_{n,1}}{l_{12}}
\end{array}
\eea
where we used
\bea
\frac{c_{n,1}}{\mu_1} +\frac{c_{n,2}}{\mu_2} &=&   \frac{v_{n,1}'(0)}{l_{12}} - \frac{f_n}{\beta}
\eea
and introduced
\bea\label{defXi1Xi2}
\xi_1 =\frac{\coth \sqrt{\mu_1}}{\sqrt{\mu_1}},\quad  \xi_2 = \frac{\coth \sqrt{\mu_2}}{\sqrt{\mu_2} }\,.
\eea
From Eq.~\ref{wn01} we find
\bea\label{ExprforVn1prime0}
c_{n,1}\xi_1 + c_{n,2}\xi_2 = - \frac{v_{n,1}'(0)}{\kappa_1} + \frac{v_{n,2}'(0)}{\kappa_2} -\frac{f_n}{\beta}  + \frac{\delta_{n,1}}{l_{12}}
\eea
and with Eq.~\ref{relationv1v2} we obtain
\bea
c_{n,1}\xi_1 + c_{n,2}\xi_2 = -v_{n,1}'(0) \( \frac{1}{\kappa_1}  + \frac{l_{21}}{l_{12} \kappa_2}\)  + f_n \( \frac{1}{l_{32}\kappa_2} -\frac{1}{\beta}\) + \frac{\delta_{n,1}}{l_{12}}
\eea
By inserting $c_{n,1}$ and $c_{n,2}$ from Eq.~\ref{eqforCn1Cn2_1} we obtain
\bea
v_{n,1}'(0) ( a_1 \xi_1 +  a_2 \xi_2)  + f_n ( b_1 \xi_1 +  b_2 \xi_2 ) =  -v_{n,1}'(0) \( \frac{1}{\kappa_1}  + \frac{l_{21}}{l_{12} \kappa_2}\)  + f_n \( \frac{1}{l_{32}\kappa_2} -\frac{1}{\beta}\) + \frac{\delta_{n,1}}{l_{12}} \,.\nn
\eea
From this we finally get
\bea\label{ExprforVn1prime}
\ds v_{n,1}'(0) = \ds  f_n \frac{d}{e} + \frac{\delta_{n,1}}{l_{12} e}
\eea
with
\bea\label{paramDandE}
\begin{array}{rcl}
\ds d &=& \ds \frac{1}{l_{32}\kappa_2} -\frac{1}{\beta} -  ( b_1 \xi_1 +  b_2 \xi_2 ) \\
\ds  e &=& \ds  a_1 \xi_1 +  a_2 \xi_2  + \frac{1}{\kappa_1}  + \frac{l_{21}}{l_{12} \kappa_2}\,.
\end{array}
\eea
We get the final expressions
\bea\label{eqforCn1Cn2_2}
\begin{array}{rcl}
\ds c_{n,1} &=&  \ds v_{n,1}'(0) a_1 + f_n b_1 = f_n( \frac{a_1 d}{e} + b_1) + \frac{\delta_{n,1}}{l_{12}}\frac{a_1}{ e}\\ \\
 \ds  c_{n,2} &=& \ds v_{n,1}'(0) a_2 + f_n b_2  = f_n( \frac{a_2 d}{e} + b_2) + \frac{\delta_{n,1}}{l_{12}} \frac{a_2}{ e}
\end{array}
\eea

\subsection*{Sojourn times and MFPT}
The scaled sojourn times are
\bea\label{SolScaledSojTimes}
\begin{array}{rcl}
\ds \hat \tau_{n,1}(x) &=& \ds v_{n,1}(x) + \bar {\hat \tau}_{n,1}
= v_{n,1}(x) - v_{n,1}(0) + \hat \tau_{n,1} (0) =  v_{n,1}(x) - v_{n,1}(0) + \frac{v_{n,1} '(0)}{\kappa_1}
\\ \\
\ds \hat \tau_{n,2}(x) &=& \ds v_{n,1}(x) - (v_{n,1}(x)- v_{n,2}(x)) + \bar {\hat \tau}_{n,2}
=  \hat \tau_{n,1}(x) - w_n(x) + \bar {\hat \tau}_{n,2} - \bar {\hat \tau}_{n,1}  \\
&=& \ds \hat \tau_{n,1}(x) -  w_n(x) +  \frac{v_{n,1}'(0) }{l_{12}} - \frac{\delta_{n,1}}{l_{12}}
\\ \\
\ds \hat \tau_{n,3}(x) &=&  \ds  \hat \tau_{n,2}(x)   - (v_{n,2}(x)- v_{n,3}(x) )+  \frac{\delta_{n,3}}{l_{32}}
\end{array}
\eea
where we used $\bar{\hat \tau}_{n,3}-\bar {\hat \tau}_{n,2} = \frac{\delta_{n,3}}{l_{32}}$ obtained by integrating Eq.~\ref{eqSojournTimes2}. The sojourn times with uniform initial distributions are
\bea
\ds \bar {\hat \tau}_{n,1} = \ds - v_{n,1}(0) + \frac{v_{n,1} '(0)}{\kappa_1}\,, \quad
\ds \bar {\hat \tau}_{n,2} = \ds  \bar {\hat \tau}_{n,1}  + \frac{v_{n,1}'(0) }{l_{12}} - \frac{\delta_{n,1}}{l_{12}}\,, \quad
\ds \bar{\hat \tau}_{n,3} = \ds\bar {\hat \tau}_{n,2}+ \frac{\delta_{n,3}}{l_{32}}  \label{meanScaledSojournTimes} \\
\bar \tau_{n,1}  =
-\frac{v_{n,1}(0)}{\nu_n}  + \frac{v_{n,1} '(0)}{\nu_n \kappa_1} \,, \quad
\ds \bar \tau_{n,2} = \ds  \bar \tau_{n,1}  + \frac{v_{n,1}'(0) }{\nu_n l_{12}} - \frac{\delta_{n,1}}{k_{12}}\,, \quad
\ds \bar\tau_{n,3} = \ds\bar \tau_{n,2}+ \frac{\delta_{n,3}}{k_{32}} \,. \label{meanSojournTimes}
\eea
The MFPT with uniform initial distribution in state $m$ is
\bea\label{scaling relationSojTimes}
\bar \tau(m) = \sum_{n=1}^3 \nu_n^{-1}  \bar {\hat \tau}_{n,m}= \sum_{n=1}^3 \bar \tau_{n,m}\,.
\eea
Because switching between states is fast compared to the overall search time, the mean sojourn times are almost independent on the initial state, $\bar {\hat \tau}_{n,1} \approx \bar {\hat \tau}_{n,2}\approx \bar {\hat \tau}_{n,3}$. By noting that $\frac{\bar {\hat \tau}_{i,m}}{f_i}\approx\frac{\bar {\hat \tau}_{j,m}}{f_j} $ we find
\bea\label{sojTimesRelations}
\frac{\bar \tau_{i,m}}{\bar \tau_{j,m} } \approx \frac{\nu_j}{\nu_i} \frac{f_i}{f_j}\,.
\eea
The expression for the MFPT simplifies to
\bea\label{mfptApprox}
\begin{array}{rcl}
\ds \bar \tau &\approx& \ds \bar \tau_{1,1}+ \bar \tau_{2,1} + \bar \tau_{3,1} = \bar \tau_{1,1}\( 1+ \frac{k_{12}}{k_{21}} + \frac{k_{12} k_{23}}{k_{21}k_{32}} \)  \\
&=& \ds N_{12} \( \frac{1}{k_{12}}+ \frac{1}{k_{21}} + \frac{k_{23}}{k_{21}} \frac{1}{k_{32}} \)
= N_{23} \( \frac{k_{21}}{k_{23}}\frac{1}{k_{12}}+ \frac{1}{k_{23}} + \frac{1}{k_{32}} \)
\end{array}
\eea
where we introduced the mean number of switchings between state  1 and 2 resp. 2 and 3
\bea
N_{12} =  \bar \tau_{1,1} k_{12}= \bar {\hat \tau}_{1,1}  l_{12}\,, \quad  \frac{N_{23}}{N_{12}} =  \frac{k_{23}}{k_{21}} \
\eea
$N_{12}$ and $N_{23}$ can be expressed as a function of the DNA length $L$ and the mean square displacements $\sigma_1$, $\sigma_2$ and $\sigma_3$.

\subsection*{Dependence of the search time on the DNA length}
To analyze how the MFPT depends on the DNA length $L$, we introduce $\hat L=\frac{L}{L_0}$, where $L_0=1bp$ is a reference length, and extract the dependency on $\hat L$. We therefore replace $l_{ij}= \frac{L^2 k_{ij}}{D_i}$ with $\hat L^2 l_{ij}$, where $l_{ij}= \frac{L_0^2k_{ij}}{D_i}$ is now evaluated with $L_0$. We proceed similarly with all the other parameters: $\nu_i \to \hat L^{-2} \nu_i$, $\kappa_i \to \hat L \kappa_i$, $\mu_i \to \hat L^2 \mu_i$, $\alpha \to \hat L^2 \alpha$, $\beta \to \hat L^4 \beta$, $f_n\to \hat L^2 f_n$, $a_i \to a_i$ and $b_i \to \hat L^{-2} b_i$. In terms of the rescaled parameters we get
\bea\label{scaledfunctionsVandW}
\begin{array}{rcl}
\ds v_{n,1}(x) &=& \ds c_{n,1} \frac{l_{12}}{\mu_1 } \( \frac{1}{\hat L^2 \mu_1} - \frac{\cosh(\sqrt{\hat L^2\mu_1} (1-x))}{\hat L \sqrt{\mu_1} \sinh \sqrt{\hat L^2\mu_1}}  \) +
  \ds c_{n,2} \frac{l_{12}}{\mu_2 } \( \frac{1}{\hat L^2\mu_2} - \frac{\cosh(\sqrt{\hat L^2\mu_2} (1-x))}{\hat L\sqrt{\mu_2} \sinh\sqrt{\hat L^2\mu_2}}  \) \\
&&  \ds - \frac{l_{12}f_n}{\beta} g(x)
\end{array}
\eea
The parameters $c_{n,1}$ and $c_{n,2}$ from Eq.~\ref{eqforCn1Cn2_2} are
\bea\label{eqforCn1Cn2_scaled}
\begin{array}{rcl}
\ds c_{n,1} &=&  \ds f_n( \frac{a_1 d}{e} + b_1) + \frac{\delta_{n,1}}{\hat L} \frac{a_1}{ l_{12}e}
\\
\ds  c_{n,2} &=& \ds  f_n( \frac{a_2 d}{e} + b_2) + \frac{\delta_{n,1}}{\hat L} \frac{a_2}{ l_{12} e}
\end{array}
\eea
with
\bea\label{paramDandE_scaled}
\begin{array}{rcl}
\ds d &=& \ds \frac{1}{l_{32}\kappa_2} -\frac{1}{\hat L \beta} -  ( b_1 \xi_1 +  b_2 \xi_2 ) \\
\ds  e &=& \ds  a_1 \xi_1 +  a_2 \xi_2  + \frac{1}{\kappa_1}  + \frac{l_{21}}{l_{12} \kappa_2}\,.
\end{array}
\eea
and
\bea
\xi_1=\frac{\coth \sqrt{\hat L^2 \mu_1}}{\sqrt{\mu_1}},\quad  \xi_2= \frac{\coth \sqrt{\hat L^2 \mu_2}}{\sqrt{\mu_2} }\,.
\eea
For a long DNA with $\hat L^2\mu_i \gg 1$ we have
\bea\label{largeLV}
v_{n,1}(0) &\approx& -\frac{l_{12}}{\hat L} \( \frac{c_{n,1}}{\mu_1 \sqrt{\mu_1} }
+   \frac{c_{n,2}}{\mu_2 \sqrt{\mu_2}} \)  - \frac{f_n l_{12}}{3\beta}\,. \\
v_{n,1}'(0) &=&  \frac{c_{n,1}l_{12}}{\mu_1} + \frac{c_{n,2}l_{12}}{\mu_2}  + \frac{f_n l_{12}}{\beta}
\eea
and from this we find for the sojourn times
\bea\label{eqSojTimeLargeL}
\bar \tau_{n,1} &=& -\frac{\hat L^2}{\nu_n}v_{n,1}(0)  + \frac{\hat L}{\nu_n \kappa_1} v_{n,1} '(0)  \nn\\
&\approx& \frac{\hat L l_{12}}{\nu_n} \(\frac{ c_{n,1} }{\mu_1 \sqrt{\mu_1} }
+  \frac{ c_{n,2} }{\mu_2 \sqrt{\mu_2}}\) + \frac{\hat L l_{12}}{\nu_n \kappa_1}  \( \frac{c_{n,1}}{\mu_1} + \frac{c_{n,2}}{\mu_2}  \)  + \frac{l_{12}f_n}{\beta} \( \frac{L^2}{3 D_n}  +  \frac{L}{\chi_1}\)
\eea
For $k_{12}=0$ ($l_{12}=0$) we recover  $\bar \tau_{1,1} = \frac{L^2}{3D_1}+\frac{L}{\chi_1}$. The number of switchings with  $\chi_1=\infty$ are
\bea\label{numberOFSwitchingsN12}
N_{12}  = \bar \tau_{1,1} k_{12} \approx \hat L l_{12}^2 \(\frac{ c_{1,1} }{\mu_1 \sqrt{\mu_1} }
+  \frac{ c_{1,2} }{\mu_2 \sqrt{\mu_2}}\) + \hat L^2 l_{12} \frac{l_{21}l_{32}}{3\beta}
\eea

\section*{Probability to bind to the target before detaching}
When a TF reaches the target in state 2 it eventually binds with probability $P$ or it detaches with probability $Q=1-P$. To compute $Q$ we consider a switching process with $k_{32}=0$ ($l_{32}=0$) to avoid rebinding to the DNA. The mean probability to detach before binding to the target when initially at the target site in state 2 is $Q = k_{23} \tau_{2,2}(0) =  l_{23} \hat \tau_{2,2}(0)$. However, we cannot use $\hat \tau_{2,2}(0)$ from our previous analysis  because state 3 is missing and $v_{n,3}(x)=0$. More specifically, Eq.~\ref{relationBetweenSojTimes} is not valid when  $v_{n,3}(x)=l_{32}=0$, and we cannot apply  Eq.~\ref{relationBetweenSojTimesPrime} to obtain $v_{2,2}'(0)$ as a function of $v_{2,1}'(0)$. We therefore recalculate $v_{2,1}'(0)$ and $v_{2,2}'(0)$ without Eq.~\ref{relationBetweenSojTimesPrime} for $l_{32}=0$ and the modified boundary condition $\hat \tau_{2,2}(0) =\frac{Q}{l_{23}}$. The parameters $\alpha$, $\beta$, $\mu_1$ and $\mu_2$ are evaluated with $l_{32}=0$, e.g. $\beta = l_{12} l_{23}$ and $\alpha= l_{12} + l_{21}+l_{23}$. By integrating Eq.~\ref{eqSojournTimes2} we find
\bea\label{eqSojournTimesIntegrated1}
\begin{array}{rcl}
\ds v_{2,1}'(0) &=& \ds -l_{12} ( \bar {\hat \tau}_{2,1} - \bar {\hat \tau}_{2,2} ) \\
\ds v_{2,2}'(0) &=&  \ds 1 +  l_{21} ( \bar {\hat \tau}_{2,1} - \bar {\hat \tau}_{2,2} ) - l_{23}  \bar {\hat \tau}_{2,2} =
1 +  ( l_{21} + l_{23}) ( \bar {\hat \tau}_{2,1} - \bar {\hat \tau}_{2,2} ) -  l_{23}  \bar {\hat \tau}_{2,1}
\end{array}
\eea
and from this we get
\bea\label{eqSojournTimesIntegrated2}
\begin{array}{rcl}
\ds \bar {\hat \tau}_{2,1} &=& \ds -\frac{1}{l_{23}} \( v_{2,2}'(0) +\frac{l_{21}+ l_{23}}{l_{12}}  v_{2,1}'(0) -1 \) \\
\ds \bar {\hat \tau}_{2,2} &=& \ds -\frac{1}{l_{23}} \( v_{2,2}'(0) +\frac{l_{21}}{l_{12}} v_{2,1}'(0) -1 \)  \,. \\
\end{array}
\eea
Eq.~\ref{ExprforVn1prime0} reads ($f_2=0$)
\bea
c_{2,1}\xi_1 + c_{2,2}\xi_2 = - \frac{v_{2,1}'(0)}{\kappa_1} + \frac{v_{2,2}'(0)}{\kappa_2} = - \frac{v_{2,1}'(0)}{\kappa_1} + \frac{Q}{l_{23}}
\eea
where we used $\frac{v_{2,2}'(0)}{\kappa_2} = \hat \tau_{2,2}(0) =\frac{Q}{l_{23}}$. From
\bea
\frac{\beta}{l_{12}}v_{2,1}(0) &=&- c_{2,1}\mu_2 \xi_1- c_{2,2}\mu_1 \xi_2 + c_{2,1} \frac{\mu_2}{\mu_1} + c_{2,2} \frac{\mu_1}{\mu_2} \nn\\
&=&- c_{2,1}\mu_2 \xi_1- c_{2,2}\mu_1 \xi_2  + \frac{v_{2,1}'(0)}{l_{12}} \alpha - w_{2}'(0)  \nn\\
&=&- c_{2,1}\mu_2 \xi_1- c_{2,2}\mu_1 \xi_2  + \frac{l_{21} + l_{23}}{l_{12}} v_{2,1}'(0) + v_{2,2}'(0) \nn\\
\frac{\beta}{l_{12}}v_{2,1}(0) &=&  l_{23}\hat \tau_{2,1}(0)- l_{23} \bar {\hat \tau}_{2,1} = \( \frac{ l_{23}}{\kappa_1} +\frac{l_{21}+ l_{23}}{l_{12}} \) v_{2,1}'(0) + v_{2,2}'(0) -1 \nn
\eea
we get
\bea
c_{2,1}\mu_2 \xi_1 +  c_{2,2}\mu_1 \xi_2 = 1-  \frac{ l_{23}}{\kappa_1} v_{2,1}'(0)
\eea
Thus, we find the system of equations
\bea
\begin{array}{rcl}
\ds c_{2,1}\mu_2 \xi_1 +  c_{2,2}\mu_1 \xi_2  + \frac{ l_{23}}{\kappa_1} v_{2,1}'(0)&=& \ds  1\\
\ds c_{2,1}\xi_1 + c_{2,2}\xi_2 + \frac{v_{2,1}'(0)}{\kappa_1} &=& \ds \frac{Q}{l_{23}}
\end{array}
\eea
By writing
\bea
c_{2,1}= \tilde a_1  v_{2,1}'(0) + \tilde b_1 v_{2,2}'(0) \,,\quad  c_{2,2}= \tilde a_2  v_{2,1}'(0) + \tilde b_2 v_{2,2}'(0)
\eea
with
\bea
\tilde a_1= \frac{\mu_1 -l_{23} }{\mu_1-\mu_2}, \quad \tilde  b_1 = \frac{\mu_1}{\mu_2-\mu_1}, \quad \tilde a_2= \frac{\mu_2 -l_{23}}{\mu_2-\mu_1},\quad \tilde b_2 = \frac{\mu_2}{\mu_1-\mu_2}
\eea
we obtain
\bea\label{matEq}
A\begin{pmatrix} v_{2,1}'(0) \\v_{2,2}'(0) \end{pmatrix}=  \begin{pmatrix} 1  \\ \frac{Q}{l_{23}}\end{pmatrix}
\eea
with the matrix
\bea
A_{ij} =
\begin{pmatrix}
\tilde a_1 \xi_1\mu_2 + \tilde a_2 \xi_2 \mu_1 + \frac{ l_{23}}{\kappa_1} & \tilde  b_1 \xi_1 \mu_2 + \tilde  b_2 \xi_2\mu_1  \\
\tilde a_1\xi_1 + \tilde a_2\xi_2 +\frac{1}{\kappa_1}  & \tilde b_1 \xi_1 + \tilde b_2\xi_2
\end{pmatrix}
\eea
The solution of Eq.~\ref{matEq} is
\bea
\begin{array}{rcl}
\ds v_{2,1}'(0) &=& \ds \frac{1}{\mbox{det}(A)} \( A_{22} - A_{12} \frac{Q}{l_{23}} \) \\
\ds v_{2,2}'(0) &=& \ds \frac{1}{\mbox{det}(A)} \( -A_{21} + A_{11} \frac{Q}{l_{23}} \)
\end{array}
\eea
From this we obtain  ($P=1-Q$)
\bea\label{exprKappa2andP}
\begin{array}{rcl}
\ds \kappa_2 &=& \ds \frac{\hat \tau_{2,2}'(0)}{\hat \tau_{2,2}(0)} = \frac{l_{23}}{Q} v_{2,2}'(0)  =\frac{1}{\mbox{det}(A)} \( A_{11} -A_{21} \frac{l_{23}}{Q}\) \\
\ds Q &=& \ds l_{23} \frac{A_{21}}{A_{11}- \kappa_2 {\mbox{det}(A)}} \,.
\end{array}
\eea
For example, for $k_{23}=0$ or $\chi_2=\infty$ we have $Q=0$. The maximum is obtained for $\kappa_2=0$
\bea
Q_{max} = l_{23} \frac{A_{21}}{A_{11}} \,.
\eea
$Q_{max}$ depends on the switching rates and on $\kappa_1$. For example, $Q_{max}=1$ is found for $\kappa_1=0$ (no binding in state 1) or $l_{21}=0$ (no switching to state 1).

\end{document}